\documentclass[twocolumn,letterpaper]{IEEEAerospaceCLS}  % only supports two-column, letterpaper format

\usepackage{silence}
\usepackage[colorlinks=true, urlcolor=blue, linkcolor=black, citecolor=black]{hyperref} % ED 2025 added for web links
\usepackage{graphicx}    % We use this package in this document
\usepackage{cite}
\usepackage{amsmath,amssymb,amsfonts}
\usepackage{algorithmic}
\usepackage{textcomp}
\usepackage{xcolor}
\usepackage{cleveref}
\usepackage{acronym}
\usepackage{subcaption}

\begin{document}

\title{A Practical Framework for Incremental Testing and Technology Maturation of Spacecraft Autonomy}

\author{%
Kerianne Hobbs\\
The Aerospace Corporation\\
14745 Lee Rd\\
Chantilly, VA 20151\\
kerianne.hobbs@aero.org
\and
Ben Bycroft\\ 
The Aerospace Corporation\\
2310 E. El Segundo Blvd.\\
El Segundo, CA 90245\\
benjamen.p.bycroft@aero.org
\and
Pamela Wood\\
The Aerospace Corporation\\
14745 Lee Rd\\
Chantilly, VA 20151\\
pamela.s.wood@aero.org
\and
Mathew W. Bissonnette\\
The Aerospace Corporation\\
14745 Lee Rd\\
Chantilly, VA 20151\\
mathew.w.bissonnette@aero.org
\and
M.J. Durst\\
The Aerospace Corporation\\
14745 Lee Rd\\
Chantilly, VA 20151\\
michael.j.durst@aero.org
\and 
Kimberly Sover\\
The Aerospace Corporation\\
2310 E. El Segundo Blvd.\\
El Segundo, CA 90245\\
kimberly.sover@aero.org
\and
Alexandra Mangel\\
The Aerospace Corporation\\
14745 Lee Rd\\
Chantilly, VA 20151\\
alex.mangel@aero.org
\and Gabriel Behrendt\\
The Aerospace Corporation\\
14745 Lee Rd\\
Chantilly, VA 20151\\
gabriel.j.behrendt@aero.org
\and Upneet Singh\\
The Aerospace Corporation\\
14745 Lee Rd\\
Chantilly, VA 20151\\
upneet.singh@aero.org
\and 
Juan Rodriguez\\
The Aerospace Corporation\\
2310 E. El Segundo Blvd.\\
El Segundo, CA 90245\\
juan.p.rodriguez@aero.org
\and
Lukas Dillingham\\
The Aerospace Corporation\\
2310 E. El Segundo Blvd.\\
El Segundo, CA 90245\\
lukas.dillingham@aero.org
\and Zeve Akerling\\
The Aerospace Corporation\\
2310 E. El Segundo Blvd.\\
El Segundo, CA 90245\\
zeve.akerling@aero.org
\and Nicholas Oune\\
The Aerospace Corporation\\
2310 E. El Segundo Blvd.\\
El Segundo, CA 90245\\
nicholas.a.oune@aero.org
\and Daniel Thomlinson\\
The Aerospace Corporation\\
2310 E. El Segundo Blvd.\\
El Segundo, CA 90245\\
daniel.thomlinson@aero.org
\and 
Evan N. Sperber\\
The Aerospace Corporation\\
2310 E. El Segundo Blvd.\\
El Segundo, CA 90245\\
evan.n.sperber@aero.org
\and Denise Burrell\\
The Aerospace Corporation\\
2310 E. El Segundo Blvd.\\
El Segundo, CA 90245\\
denise.galindo@aero.org
\and Phaedrus Leeds\\
The Aerospace Corporation\\
2310 E. El Segundo Blvd.\\
El Segundo, CA 90245\\
phaedrus.leeds@aero.org
\and Brian Hardy\\
The Aerospace Corporation\\
2310 E. El Segundo Blvd.\\
El Segundo, CA 90245\\
brian.s.hardy@aero.org
\and
Nathaniel Hamilton\\
University of Dayton Research Institute\\
1700 S Patterson Blvd \\
Dayton, OH 45469\\
Nate.Hamilton@udri.udayton.edu
\and
Raul Santos\\
Texas A\&M University\\
701 Ross St\\
College Station, TX 77843\\
rdsflomo03@tamu.edu
\and
Arturo de la Barcena\\
University of Dayton Research Institute\\
1700 S Patterson Blvd \\
Dayton, OH 45469\\
arturo.delabarcena@udri.udayton.edu
\and
Kyle Dunlap\\
Air Force Research Laboratory\\
2241 Avionics Circle\\
Wright--Patterson AFB, OH, 45433\\
kyle.dunlap.5@us.af.mil
%%%% IMPORTANT: Use the correct copyright information--IEEE, Crown, or U.S. government. %%%%%
%\thanks{\footnotesize 979-8-3315-7360-7/26/$\$31.00$ \copyright2026 IEEE} 
%\thanks{\footnotesize 979-8-3195-1048-8/27/\$31.00 \copyright2027 IEEE}    % This creates the copyright info that is the correct 2026 data.
%\thanks{{U.S. Government work not protected by U.S. copyright}}         % Use this copyright notice only if you are employed by the U.S. Government.
%\thanks{{979-8-3315-7360-7/26/$\$31.00$ \copyright2026 Crown}}          % Use this copyright notice only if you are employed by a crown government (e.g., Canada, UK, Australia).
%\thanks{{979-8-3315-7360-7/26/$\$31.00$ \copyright2026 European Union}}    % Use this copyright notice is you are employed by the European Union.
}

\maketitle

\thispagestyle{plain}
\pagestyle{plain}

\begin{abstract}
This paper addresses the ``trust gap" in deploying AI-enabled spacecraft autonomy through an integrated verification framework combining offline ground testing with online flight monitoring. The offline phase progresses systematically through simulation studies, processor-in-the-loop (PIL), hardware-in-the-loop (HIL), and robotic emulation testing, culminating in flight validation. The online phase employs AI watchdog systems for performance monitoring and safety guardrails for real-time constraint enforcement. Key contributions include: (1) systematic organization of verification techniques emphasizing closed-loop behavioral testing at each phase, (2) integration of AI watchdogs with runtime assurance, and (3) demonstration of the framework through a reinforcement learning-based attitude control case study. This approach integrates mature technologies to provide a viable pathway for safe deployment of AI-enabled autonomy in high-consequence space operations while addressing simulation-to-reality transfer challenges.
\end{abstract}

\vspace{-3.3em}
\tableofcontents

\section{Introduction}

The space domain is experiencing unprecedented growth driven by large satellite constellations, commercial in-space servicing, and cislunar exploration. These developments demand increasingly autonomous spacecraft to manage operational complexity, enable rapid decision-making in proximity operations, and overcome communication latencies. Deploying AI-enabled autonomy in this high-consequence environment presents significant challenges: failures can result in mission loss, degradation of critical services (communications, navigation, Earth observation), collision-generated debris, or cascading impacts across orbital regimes. Acceptable autonomy must avoid operational hazards (collisions, fuel depletion), system failures (structural damage, thermal issues), and loss of controllability \cite{hobbs2021risk}.

The fundamental barrier to adoption is a ``trust gap": systems are insufficiently tested to be trusted, yet insufficiently trusted to receive extensive testing opportunities \cite{hobbs2023space}. Traditional spacecraft \ac{VandV}, developed for deterministic rule-based systems, is insufficient for AI-enabled autonomy. A neural network control system (NNCS) trained via reinforcement learning (RL) presents unique challenges including the simulation-to-reality transfer gap, opaque decision-making processes, and unpredictable behavior in novel situations. While formal verification methods for neural networks exist, they remain limited to small state spaces and short time horizons, making them impractical for complex spacecraft control.

This paper presents an integrated verification framework combining systematic offline ground testing with online flight monitoring to enable safe maturation of AI-enabled autonomous spacecraft. The offline phase builds confidence through progressive realism: simulation studies identify failure modes, \ac{PIL} testing validates algorithm performance and timing on the flight processor, \ac{HIL} testing reveals performance issues with physical sensors and actuators, and robotic emulation enables evaluation of novel sensing paradigms and robustness testing in the presence of real-world noise. Critically, each phase emphasizes \textit{closed-loop behavioral testing} of autonomous decision-making, not merely software integration. The online phase deploys layered safety mechanisms: AI watchdog systems predict performance degradation through internal neural network monitoring, while safety guardrails enforce hard constraints via \ac{RTA}. We demonstrate this framework through a reinforcement learning attitude controller progressed from simulation to on-orbit validation, showing how systematic testing builds confidence while identifying improvements at each stage.

\subsection{Contributions}

This paper systematically integrates established verification techniques into a cohesive framework specifically for AI-enabled spacecraft autonomy. While individual testing methods are mature, their systematic combination with emphasis on closed-loop behavioral testing represents a novel yet practical approach to addressing the trust gap. Key contributions include:

\begin{itemize}
    \item Systematic integration of complementary offline and online verification techniques showing progression through \ac{TRLs};
    \item Emphasis on closed-loop behavioral testing at every phase to explicitly evaluate autonomous decision-making, not merely software integration;
    \item Integration of AI watchdog systems with \ac{RTA} guardrails for complementary monitoring and intervention during flight operations; and 
    \item Case study demonstrating framework progression from simulation to on-orbit validation for a \ac{RL}-based neural network attitude controller.
\end{itemize}

The remainder of this paper is organized as follows: Section~\ref{sec:related} reviews background information; Sections~\ref{sec:offline} and~\ref{sec:online} detail the offline and online verification phases of the framework; Section~\ref{sec:casestudy} presents the case study; and Section~\ref{sec:conclusions} concludes with lessons learned.

\section{Background}\label{sec:related}
Traditional spacecraft \ac{VandV}, developed for deterministic rule-based systems, emphasizes software integration (i.e., verifying that code executes correctly on flight hardware) rather than evaluating autonomous decision-making behavior. Testing often uses pre-recorded data from a simulation rather than closed-loop simulation, preventing evaluation of how autonomous decisions affect vehicle state evolution over time. This approach is insufficient for AI-enabled autonomy, where behavior depends not only on the control algorithm itself but also on timing, interactions within middleware frameworks such as \ac{ROS} \cite{macenski2022robot}, and the actual response of sensors and actuators to commands. Recent work on verification of learning-enabled aerospace systems has explored formal methods \cite{mandal2024safe, arjomandbigdeli2024verification} and various in-the-loop testing approaches, but these remain limited to small state spaces and short time horizons. No comprehensive framework has emerged that systematically integrates progressive offline testing with online monitoring specifically for AI-enabled spacecraft autonomy.

\subsection{Reinforcement Learning for Spacecraft Control}

\begin{figure}[htbp]
\centerline{\includegraphics[width=0.7\columnwidth]{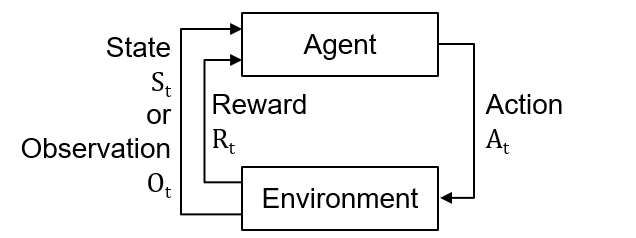}}
\caption{The agent outputs an action that updates the environment, producing a new state or observation for the next action. During training, the agent receives rewards to improve its action selection.}
\label{fig:rl_loop}
\end{figure}

After achieving performance that defeats human grandmasters in games with high-dimensional state spaces (e.g., Go \cite{silver2016mastering}) and complex strategy (e.g., StarCraft \cite{vinyals2019grandmaster}), \ac{RL} \cite{sutton1998reinforcement, sutton1999reinforcement} has emerged as a promising approach for robotics and vehicle control \cite{tang2025deep}. \ac{RL} agents have been used for aircraft flight control \cite{DeMay2022} and spacecraft attitude control \cite{gist2025}. These \ac{RL} agents learn optimal policies through trial-and-error over the course of many simulations on a ground-based high-performance computer, using algorithms like \ac{PPO} \cite{schulman2017proximal}. The community has converged on standard interfaces such as OpenAI Gymnasium \cite{brockman2016openai} and PettingZoo \cite{terry2021pettingzoo}, as well as algorithm libraries like Stable Baselines \cite{JMLR:v22:20-1364} and Ray RLLib \cite{ray_team_rllib,liang2018rllib}, facilitating development and comparison. A trained \ac{NNCS}, which is a subcategory of an \ac{RL} agent designed specifically for controlling a physical system like a robot or vehicle, finds the optimal action to take given the current observation of the state of the system, guided by a reward function intended to describe the ideal behavior. Like control theory, RL-based control takes place in a loop that evolves with time, as depicted in Figure~\ref{fig:rl_loop}.

However, \ac{RL} presents unique verification challenges that motivate our framework approach. While RL algorithms and interfaces for training in simulation environments have converged to standards, the level of fidelity of the simulation used to train algorithms remains an open research question. Because the training of an RL-based autonomous controller requires thousands to millions of simulations, there is a balance that must be achieved where the simulation should be simple enough to be computationally efficient but detailed enough to produce relevant results that will achieve the desired performance. Some concepts include curriculum learning \cite{narvekar2020curriculum} where initial training begins on an academic, textbook-level, first-principles model of a robotic system, and then the training is refined on a higher fidelity model.

\subsection{Training Versus Inference}

While training is time-consuming and computationally expensive, \textit{inference} (executing a trained \ac{AI} agent) is fast and efficient. The trained neural network control system is typically static and deterministic, outputting the same action for identical observations. Additionally, \ac{RL}-trained networks are usually simple (e.g., multi-layer perceptrons with 2-6 hidden layers of 32-512 nodes each), making them well-suited for computationally constrained environments such as embedded computers on robots and spacecraft.

\subsection{The Simulation-to-Reality Transfer Challenge}

The simulation-to-reality (sim-to-real) transfer gap represents a critical challenge for \ac{RL}-based control. Policies trained in simplified simulations may fail in real environments due to unmodeled dynamics, sensor noise, or actuator limitations. Because simulations cannot perfectly represent reality and must balance fidelity with computational efficiency, an iterative refinement process is often required: failed transfers identify missing simulation elements, prompting retraining with higher fidelity models.

In terrestrial robotics, this iterative approach through trial-and-error may be acceptable. However, the space environment is significantly less forgiving of mistakes and demands more systematic risk mitigation. This fundamental challenge motivates our framework's progressive offline testing phases that incrementally introduce real-world elements (flight processors, physical hardware, environmental disturbances) before flight deployment, combined with online monitoring and intervention capabilities during operations.

\subsection{Control Architecture Context}

Autonomy functions generally fall into three categories \cite{Hobbs2025Autonomy, fong2018autonomous}: sensing and perception, reasoning and acting, and collaboration and interaction. This paper specifically focuses on \ac{VandV} of \ac{AI}-enabled autonomous reasoning and acting, while verification approaches for the other autonomy functions are left to future work. 

\begin{figure}[htbp]
\centerline{\includegraphics[width=\columnwidth]{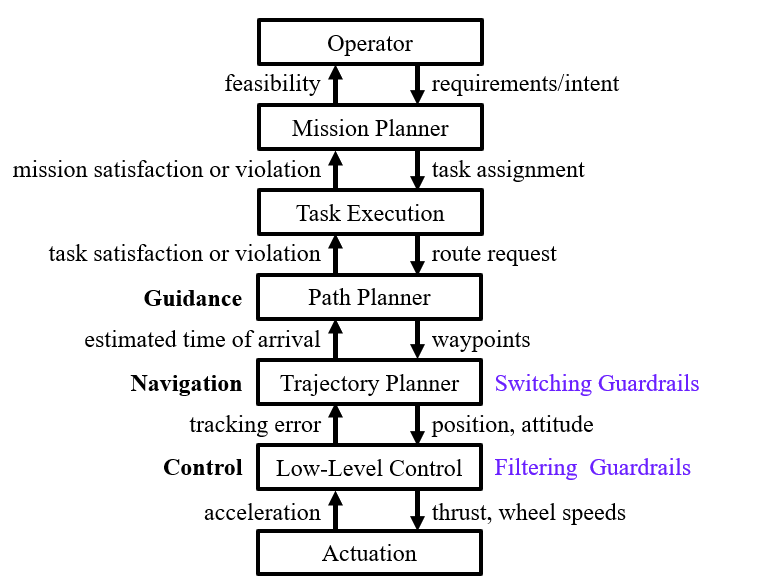}}
\caption{A notional layered control architecture that translates operator requirements and intent to vehicle actuation.}
\label{fig:LayeredControlHobbs}
\end{figure}

Understanding where autonomous reasoning and acting takes place is essential for designing appropriate offline and online verification methods. There are many ways to describe layered control architectures \cite{beard2012small,matni2024quantitative}. A traditional guidance, navigation, and control loop might look like that shown in Figure~\ref{fig:LayeredControlHobbs}. At the top layer, a human operator expresses requirements or intent to a mission manager. Next, a mission planner (which is traditionally human but may be automated or autonomous) decides which tasks the system needs to complete to satisfy the mission. A task execution layer then generates route requests to a guidance or path planner layer that commands waypoints or other path primitives. These waypoints are used in the trajectory planning layer to output control commands like position or attitude, which are then sent to a low-level control layer that outputs commands like thrust or \ac{RWA} speeds to the actuation layer. Sensing and perception elements communicate vehicle translational or rotational acceleration, tracking error, estimated time of arrival, and task tracking to evaluate mission satisfaction or violations to the mission planner, which reports on feasibility of successfully completing the mission to the operator.

\begin{figure*}[t]
\centerline{\includegraphics[width=0.9\textwidth]{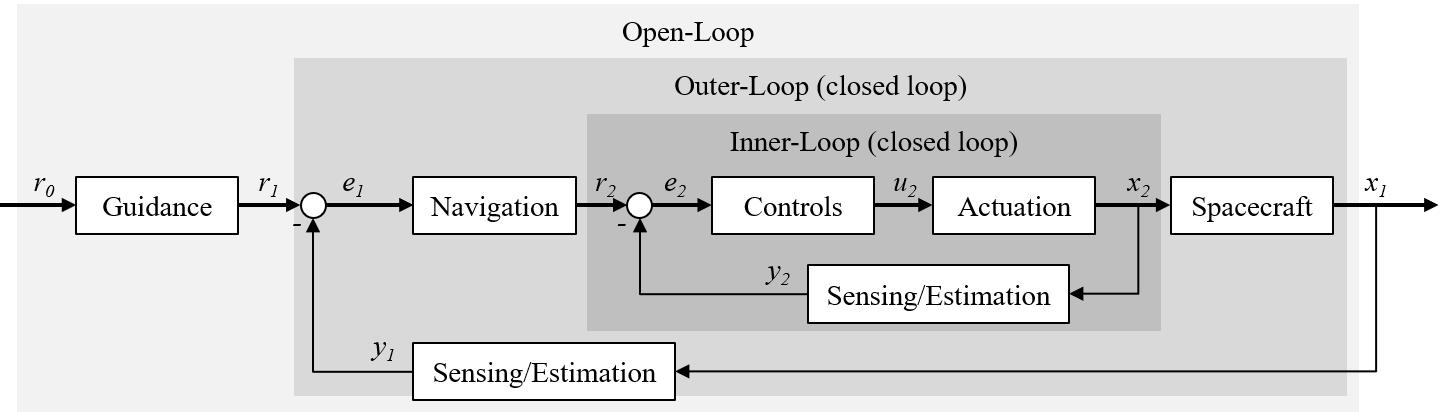}}
\caption{Notional multi-loop control system architecture where the reference input $r$ represents a desired outcome, $x$ represents the actual state, $y$ is an estimate of the state from sensor measurements, and $e$ is the error between the desired reference command and the state achieved.}
\label{fig:multi_loop}
\end{figure*}

In the context of understanding the importance of a verification framework that emphasizes closed loop behavioral testing, it is important to understand that autonomous decisions and actions take place in loops that evolve forward in time, as shown in Figure~\ref{fig:multi_loop}. Each loop uses feedback on its performance to adjust the commands at the next timestep. Over time, layers of the control loop work together to converge on the desired outcome. The lowest level control is sometimes referred to as inner-loop control, while the next highest layer may be referred to as the outer-loop, and a control layer that does not use feedback on the current state of the vehicle may be referred to as open-loop. 

\subsection{Examples of \ac{AI}-Enabled Spacecraft Control}

Early experiments in spacecraft control have focused on autonomous attitude control, likely because it is simpler to train (three degrees of freedom vs. six) and safer to test. Attitude control poses some risk to power generation (solar panel pointing) and communication (antenna pointing), but these are manageable compared to translational control risks. The University of Würzburg demonstrated autonomous attitude control using \ac{RL}-trained neural networks for low-level control \cite{gist2025}, optimizing commands to reaction wheels and magnetorquers. \ac{AFRL}'s \ac{STARS} program \cite{hobbs2025STARS} conducted extensive \ac{RL} research for spacecraft control, which was later adapted for waypoint-level autonomous control \cite{santos2026deep}, and serves as the flight test examined in this paper's case study. 

\begin{figure*}[t]
\centerline{\includegraphics[width=\textwidth]{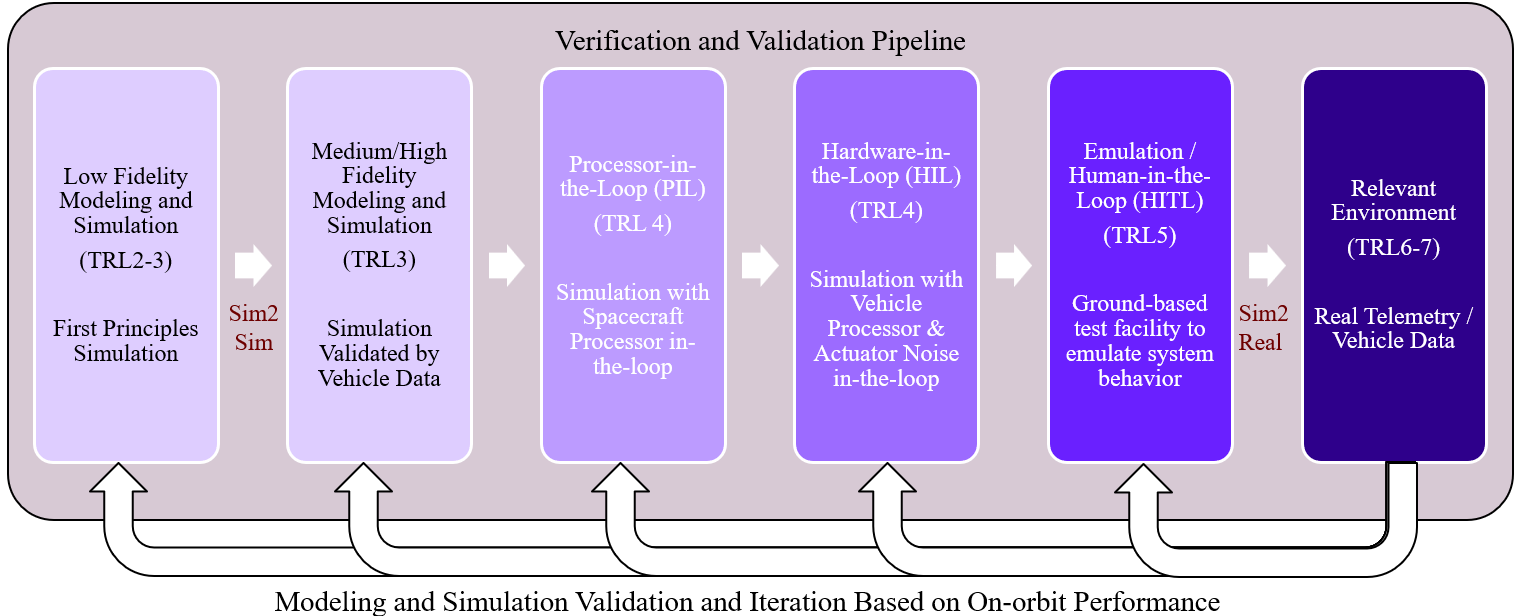}}
\caption{Offline verification phase progression flowchart showing the iterative refinement process from low fidelity simulation, through high fidelity simulation, processor-in-the-loop, hardware-in-the-loop, and emulation test, culminating in on-orbit test.}
\label{fig:offline_phases}
\end{figure*}

\section{Offline Testing Phase} \label{sec:offline}

This paper focuses on mature, practical approaches that can be implemented today to both advance and evaluate \ac{AI}-enabled autonomous reasoning and acting for spacecraft. This section focuses on offline techniques that can be completed on the ground prior to deployment. The offline testing phase progresses through increasing levels of realism and integration, as depicted in Figure~\ref{fig:offline_phases}. Each phase builds upon the previous one, with defined qualification criteria for advancement.

\subsection{Simulation Studies}

In this paper, simulation studies evaluate autonomous performance in closed loop with a spacecraft model, at various levels of fidelity. These studies are conducted entirely in software on a computer or high‑performance computing system, without any physical spacecraft components.

\subsubsection{Lower Fidelity Simulation}

Lower-fidelity simulations, sometimes also called \ac{FP} simulations when they refer to textbook-level dynamics equations, are used for initial training and Monte Carlo-style assessments of performance, repeatability, and robustness. This phase enables corner- and edge-case identification via search methods or emerging testing approaches such as adaptive stress testing when computational cost is low \cite{behrendt2026combinatorial}. Performance analysis conducted very early in the design process serves as a quick sanity check. If the algorithm performs poorly in low-fidelity simulation, the results help identify where to improve the design or training process. This phase also supports testing and refinement of metrics, success criteria, and stakeholder expectations. The algorithm does not need to be ``perfect'' at this stage, but it should satisfy basic qualification gates before investing resources in higher-fidelity testing.

\subsubsection{Higher Fidelity Simulation}

Higher-fidelity physics-based simulations use more realistic and detailed models that are validated against spacecraft data. Additional subsystems are represented with greater detail, creating a ``digital twin" that more accurately predicts vehicle behavior. This phase begins to reveal integration challenges and unexpected subsystem interactions, for example when actuator saturation and other nonlinear effects start to influence performance.

\subsection{Processor-in-the-Loop (PIL) Testing}

\begin{figure}[htbp]
\centerline{\includegraphics[width=0.6\columnwidth]{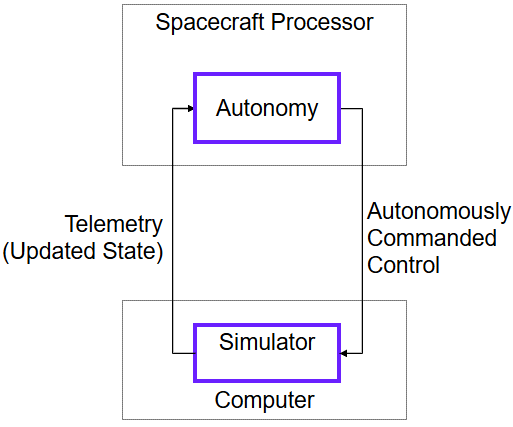}}
\caption{Processor-in-the-loop testing architecture showing the closed-loop connection between flight processor and simulator.}
\label{fig:pil}
\end{figure}

A key element of \ac{PIL} testing is the ``in-the-loop'' aspect. Traditional processor testing may only test that the system functions given an input; it is a test of integration of the software and may be used to measure computational resource use, timing, and throughput, but not a performance test of the autonomy behavior \cite{dunlap2025space}. Testing performance of autonomous behavior requires connection to a simulator in the loop to see how its decisions evolve over time. The autonomy runs on the flight processor connected to the simulator, as shown in Figure~\ref{fig:pil}. The simulator provides an observation in the right format (e.g., telemetry messages) to the processor, the autonomy on the processor outputs an action, and that action is propagated through the simulation. In summary, the processor component tests how the autonomy will be integrated on the vehicle, while the connection to the high fidelity simulation tests the autonomy's behavior. 

\begin{figure*}[t]
\centerline{\includegraphics[width=0.7\textwidth]{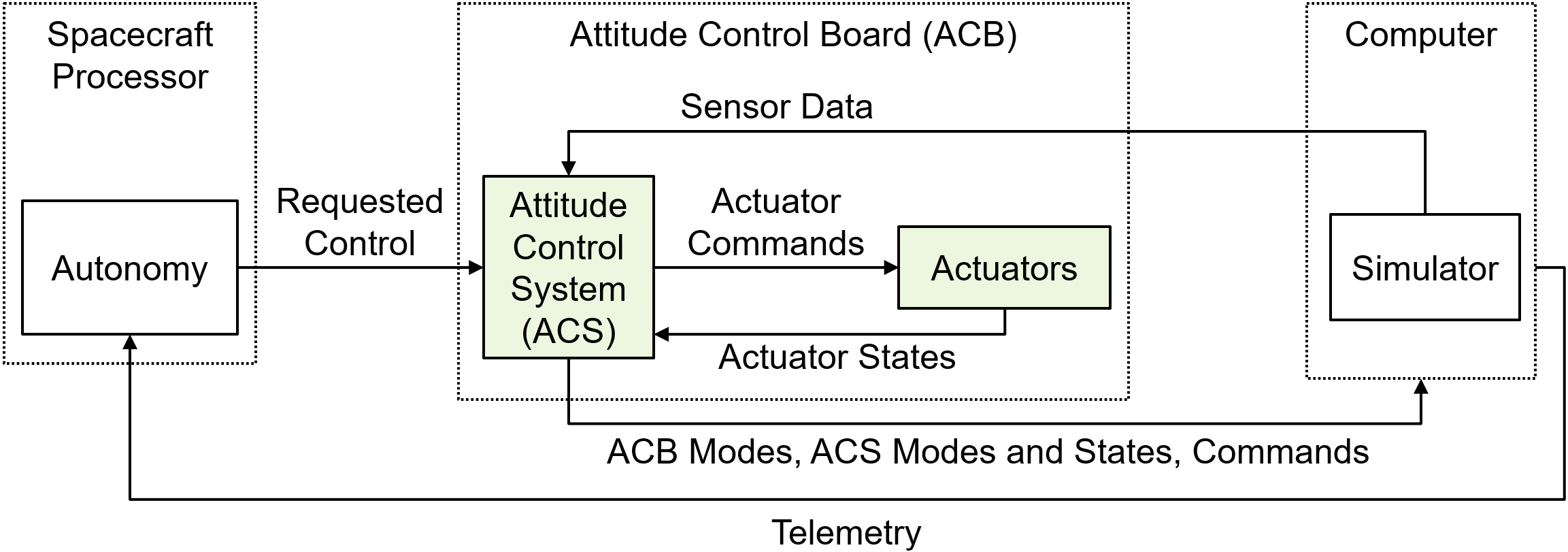}}
\caption{Hardware-in-the-loop testing architecture showing the closed-loop connection between flight processor, simulator, and attitude control system hardware.}
\label{fig:hil}
\end{figure*}

\subsection{Hardware-in-the-Loop (HIL) Testing}

\ac{HIL} expands upon the \ac{PIL} foundation by integrating flight hardware such as an \ac{ACB}, actuators, and sensors. This testing phase verifies proper integration of the sensor, actuator, and control board interfaces, validates autonomous behavior when connected to actual (rather than simulated) sensors and actuators, and evaluates the closed-loop performance of the autonomous system at a higher level of fidelity with spacecraft components. The \ac{HIL} configuration used in this work, depicted in Figure~\ref{fig:hil}, runs autonomy algorithms on the spacecraft processor and sends commands to the \ac{ACB}, which then drives a \ac{RWA}. Subsystem information from the \ac{ACB} is transmitted to a spacecraft simulation running on a standalone terrestrial computer, which propagates the state forward and provides updated state information back to the autonomy algorithms on the processor.
 
\ac{HIL} testing can be expanded to include injection of faults (such as faulty actuators or sensors) to evaluate the response of the autonomous system, or additional fidelity such as measuring the actual force generated by the actuators; however, neither of these has yet been implemented in this work. Additionally, hardware-in-the-loop testing may incorporate testbeds such as air bearings; however, the addition of these components begins to venture into the next section, robotic emulation testing.

\begin{figure}[htb]
\centerline{\includegraphics[width=\columnwidth]{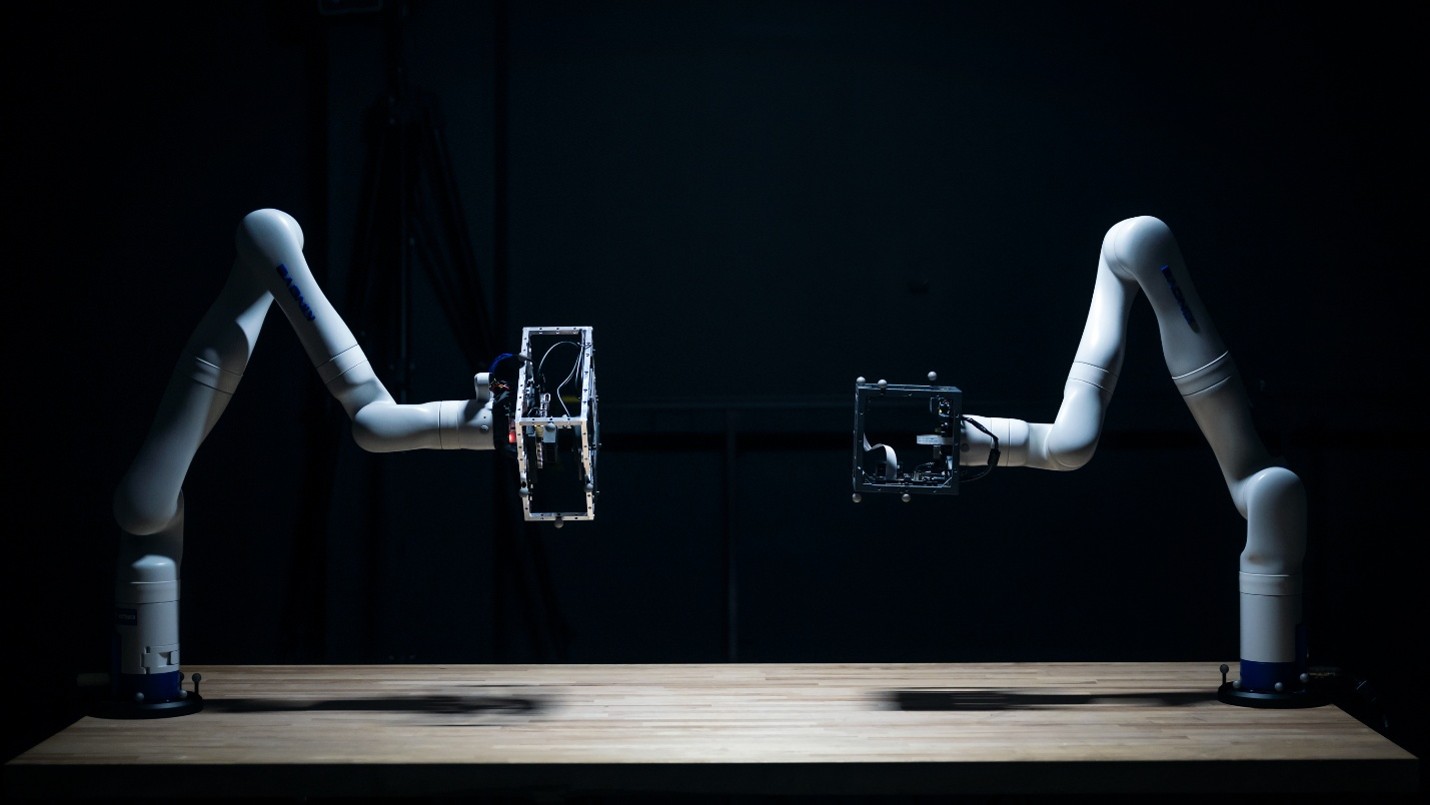}}
\caption{Two mocksats mounted to the end of robotic arms in the Aerospace Corporation’s \ac{CAVE} lab in El Segundo, CA follow tracks from a Basilisk simulation in the space environment to test a sensor’s abilities to track fiducial markers.}
\label{fig:cave}
\end{figure}

\subsection{Robotic Emulation Testing}

Robotic emulation testing uses physical components such as air bearings \cite{dcsl_astros}, robotic arms, drones \cite{phillips2024emulation, de2024testing}, neutral buoyancy facilities \cite{akin_nbrf}, granite slabs \cite{stanford_caesar_facilities}, or other low-friction surfaces \cite{chung_arcl_facilities} to emulate the space environment. Emulation is used because the robots are physically mimicking the motion of the satellite rather than simulating it numerically.

Air bearings are popular, relatively accessible platforms for evaluating spacecraft attitude dynamics. The low friction of the ball joint allows testing of a variety of attitude control actuators and components. A mockup of the entire vehicle can be mounted for testing, or components such as \ac{RWA}s, control moment gyros, or magnetorquers (particularly when the testbed is inside a Helmholtz cage that mimics a magnetic field) can be mounted to a larger attitude testbed to evaluate autonomous control capabilities.

Robotic arms and drones are typically used to follow the path of a spacecraft simulated on a computer. The spacecraft simulation is often scaled in size from a space reference frame to a laboratory reference frame (e.g., 1000 m in the space frame is 1 m in the laboratory frame) and can be scaled in time (e.g., 1 hour in the space simulation is 1 minute in the laboratory emulation). While robotic arm and drone emulation approaches provide a useful visualization of how spacecraft autonomy behaves, the primary benefits are the injection of real-world noise to evaluate autonomy algorithm robustness and the testing of sensors (especially computer vision). Robotic arms and drones never perfectly follow the commanded motion, as the physical environment introduces disturbances such as friction, air resistance, and actuation delays. When used in conjunction with advanced motion capture systems, noise from the laboratory environment is magnified in the space environment simulation (e.g., centimeters in the lab can correspond to meters in the space frame), enabling testing of autonomy robustness to significant noise. When they can support the weight of sensor payloads, these emulation platforms can also be used to test a variety of sensing and perception paradigms in support of autonomous system testing. An example of this is shown in Figure~\ref{fig:cave}.

Neutral buoyancy facilities, granite slabs, and other low-friction surfaces are particularly useful for testing close-proximity operations and contact dynamics in support of autonomy algorithms for missions such as satellite servicing, refueling, or active debris removal. Contact dynamics are particularly challenging in frictionless environments, so testbeds that emulate this environment are helpful for identifying critical design flaws in procedures or capture mechanisms, as well as in the autonomy control algorithms.

\subsection{\ac{HITL} Testing}
\ac{HITL} testing focuses on the human-system interface, interactions, and teaming. \textit{Human-autonomy interactions} are exchanges of information between humans and autonomous systems, while \textit{human-autonomy teaming} is a collaborative partnership where humans and autonomous systems share tasks, roles, workload, maintain situational awareness, and coordinate actions. An important aspect of humans interacting with autonomous systems is \textit{trust}, defined as ``the attitude that an agent will help achieve an individual’s goals in a situation characterized by uncertainty and vulnerability" \cite{lee2004trust}. Ideally trust is appropriately \textit{calibrated} to the performance of the system and the situation in which it will be used \cite{hobbs2023space, lyons2023responsible, hobbs2024safety,DTEAI2025}.

\ac{HITL} testbeds provide an opportunity for autonomous systems designers to evaluate interface and teaming concepts, including how to set safety and mission constraint boundaries on autonomous decisions and actions, measure situational awareness \cite{endsley1988situation,parasuraman2008situation,endsley2021systematic,endsley1998comparative,vidullch1991using} and workload \cite{hart2006nasa}, and design mechanisms to transfer authority for tasks and functions between humans and autonomy. 
Testing these interfaces early in prototype development enables feedback that informs the broader autonomous system design and results in more robust systems that operators will actually use. As the system matures, \ac{HITL} testbeds can also serve as training environments to familiarize operators with expected behaviors and the limits of the autonomous system.

\section{Online Verification Phase}\label{sec:online}

The online verification phase deploys layered safety mechanisms during flight operations to monitor and intervene when necessary to maintain safety and mission success.

\subsection{\ac{AI} Watchdog Systems}

Watchdog systems monitor the quality and safety of autonomous decisions in real time and alert when performance or confidence degrades. While watchdogs can be used to monitor many aspects of a system (e.g., health), this paper focuses specifically on watchdogs that assess autonomous decision-making. In particular, we consider watchdogs for autonomy that model and predict the performance of neural network-based controllers \cite{lillo2026training, bycroft2024malleable, bycroft2026operational, bycroft2024training}. Similar technology has been demonstrated for computer vision algorithms for uses like pose estimation \cite{bycroft2024capabilities}, but here we focus on anticipating degraded or unvalidated performance in the \ac{NNCS}. These watchdogs can identify when the system is operating outside its validated regime, where degraded performance is expected or watchdog predictions are less certain. This capability can be used for out-of-distribution, anomaly, novelty, or outlier detection, and for open set rejection \cite{yang2024generalized}.

\begin{figure}[htb]
\centerline{\includegraphics[width=\columnwidth]{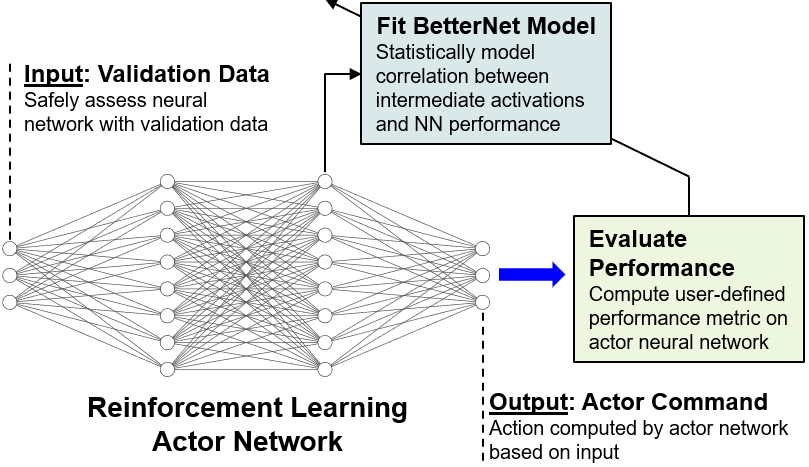}}
\caption{BetterNet Model Fit Methodology (Offline): The BetterNet model is a statistical model that is fit offline with simulation data to correlate neural network intermediate activations to a user-defined performance metric.}
\label{fig:BNetOffline}
\end{figure}

\begin{figure}[htb]
\centerline{\includegraphics[width=\columnwidth]{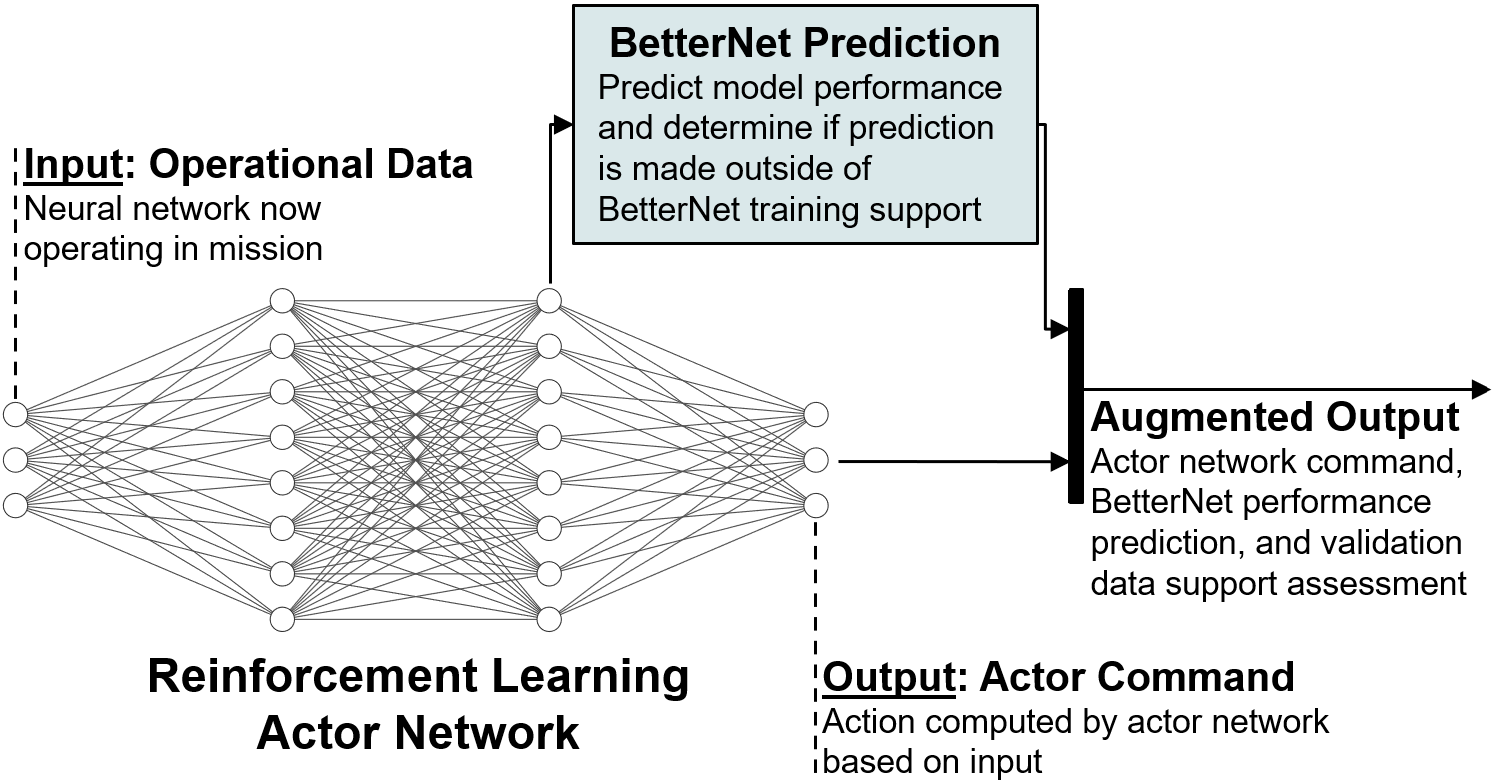}}
\caption{BetterNet Model Watchdog Application (Online): The BetterNet model is now leveraged in mission. The statistical model is used to predict performance and detect when operating outside of the training regime used to fit the BetterNet model
}
\label{fig:BNetOnline}
\end{figure}

\subsection{Safety Guardrails}

\begin{figure}[htb]
    \centering
    \includegraphics[width=\columnwidth]{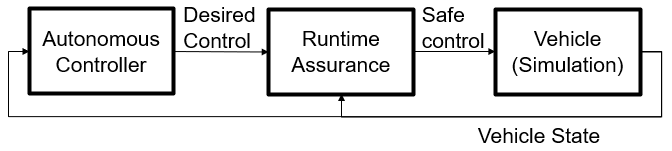}
    \caption{An autonomous closed-loop control system with an \ac{RTA} filter.}
    \label{fig:RTA}
\end{figure}

Safety guardrails, also called \ac{RTA}, are architectural solutions that intervene to prevent autonomy from violating safety or mission constraints. While a comprehensive review is available in \cite{hobbs2023runtime}, this section provides a brief overview. Autonomous decisions and actions often face state constraints (position, velocity, attitude, angular rates), resource constraints (power, fuel, thermal), and operational constraints (keep-out zones, communication windows). These constraints can be hard constraints when it is unacceptable to ever violate them, or soft constraints when violations are undesirable but permitted when otherwise unavoidable. Without some form of RTA, an autonomous controller sends a control signal (such as a thrust command) to the vehicle, which is then implemented, resulting in a new vehicle state that is sensed and estimated in a loop as depicted. \ac{RTA} places a filter between the autonomous controller and the vehicle to monitor the desired control and either pass through safe commands or modify unsafe commands to ensure adherence to safety and mission constraints, as depicted in Figure~\ref{fig:RTA}.

\subsubsection{Recovery and Reversionary Guardrails}
\textit{Recovery} guardrails safeguard against specific safety or mission violations, such as intervening to prevent a collision, but do not serve as a substitute for the autonomous controller. \textit{Reversionary} guardrail systems, on the other hand, switch from an autonomous controller to an alternate controller, such as a rule-based or human-scripted backup plan that can also complete the mission, when the behavior of the autonomous controller becomes unacceptable.

\subsubsection{Explicit vs. Implicit Monitoring}
Guardrail monitoring falls into two general categories. \textit{Explicit} monitors use mathematical expressions (e.g., inequality equations) to precisely define acceptable system states. Because it is not always possible to mathematically define all constraints, \textit{implicit} monitoring projects a trajectory of the vehicle states forward in time to predict whether it will stay within constraints.  

\subsubsection{Switching vs. Filtering Intervention}
\begin{figure}
    \centering
    \includegraphics[width=\columnwidth]{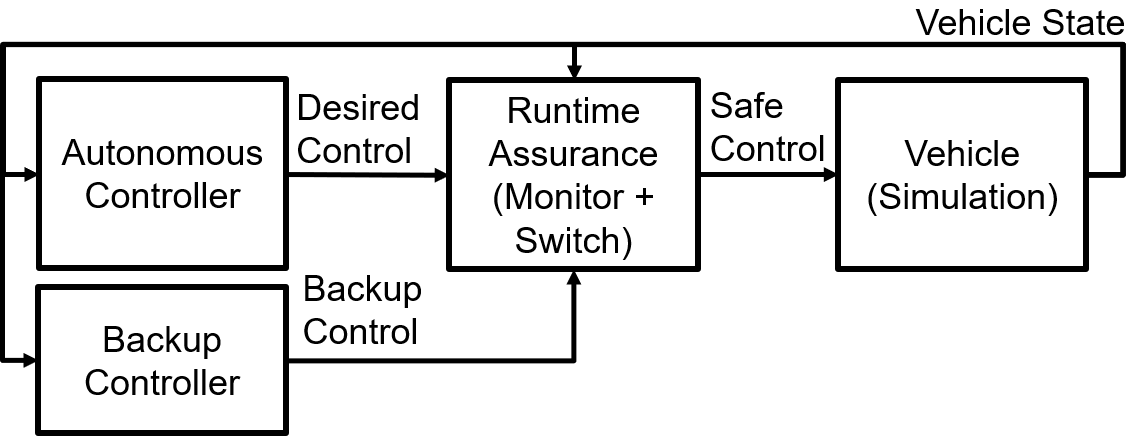}
    \caption{A switching runtime assurance architecture.}
    \label{fig:SwitchingRTA}
\end{figure}

Once a violation of safety or mission constraints is predicted, guardrails can filter the control or switch to an alternative control. \textit{Switching} guardrails, as depicted in Figure~\ref{fig:SwitchingRTA}, switch from the autonomous controller to a backup. A switching recovery guardrail might execute a predefined backup maneuver, such as a satellite entering safe mode by orienting its solar panels toward the Sun and contacting ground control, while a switching reversionary controller may switch to a less complex trusted controller to complete the mission. 
\textit{Filtering} guardrails more closely match the architecture in Figure~\ref{fig:RTA}, and modify the desired control signal rather than completely replacing it. The active set invariance filter \cite{gurriet2018online,ames2019control} is a modern technique that solves an online optimization problem to stay as close as possible to the desired control output of the neural network while satisfying control barrier function-defined safety and mission constraints. Filtering \ac{RTA} efficiently combines multiple constraints into a single framework to ensure all are met simultaneously, while switching controllers generally address one or a few constraints at a time and can be more conservative.  

\subsubsection{Command Reference Governors} 
Guardrail effectiveness depends on which level of the control architecture the system can access. As discussed earlier, autonomous control operates through a hierarchical structure (depicted in Figure~\ref{fig:LayeredControlHobbs}), where each component can be implemented in multiple ways and each layer is potentially designed by different vendors. Waypoint-level guardrails are limited to command reference governors \cite{garone2017reference} or switching-based designs. These approaches tend to be conservative (intervening when not strictly necessary) and are limited in the types of safety constraints they can directly satisfy, particularly lower-level constraints such as torque and velocity constraints. Access to lower-level control enables filtering techniques, which can simultaneously satisfy multiple safety and mission constraints while being minimally invasive to the autonomous control function.
\begin{figure}
    \centering
    \includegraphics[width=\columnwidth]{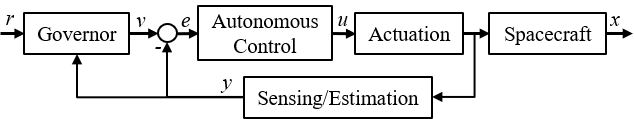}
    \caption{A command reference governor guardrail architecture that modifies the input to the autonomous controller rather than its output to assure safety.}
    \label{fig:CRG}
\end{figure}

\subsection{Integrated Watchdogs and Guardrails Across the Autonomous System Lifecycle}

Testing and fielding \ac{AI}-enabled autonomous systems in partially observable, uncertain, noisy, high-stakes environments requires assurance mechanisms beyond traditional offline verification. Integrated online assurance, including watchdogs that monitor system confidence and guardrails that bound autonomous behavior, enables safe evaluation of autonomous systems near operational limits where the most valuable performance data resides. Without mature \ac{RTA} capabilities, many high-value autonomous missions remain too risky to field. Throughout the lifecycle, watchdogs and guardrails work synergistically: during training, guardrails accelerate shielded reinforcement learning by bounding exploration to safe states; during testing, they increase efficiency by modifying rather than terminating autonomy when constraints are approached, maximizing finite test windows; and during operations, they enable graceful degradation by switching to alternative controllers when confidence is low or by applying minimally invasive filtering interventions to maintain mission effectiveness. This approach builds operator trust through transparent monitoring while avoiding ``safe mode" responses that functionally stop the mission, supporting adaptive response to both anticipated edge cases and unanticipated operational conditions.

\section{Case Study}\label{sec:casestudy}

This section will present a case study of applying the Spacecraft Autonomy \ac{VandV} Framework to a reinforcement learning-based neural network control system for spacecraft attitude.

\subsection{Representative Mission Scenario Description}

\begin{figure}[htb!]
    \centering
    \includegraphics[width=0.6\columnwidth]{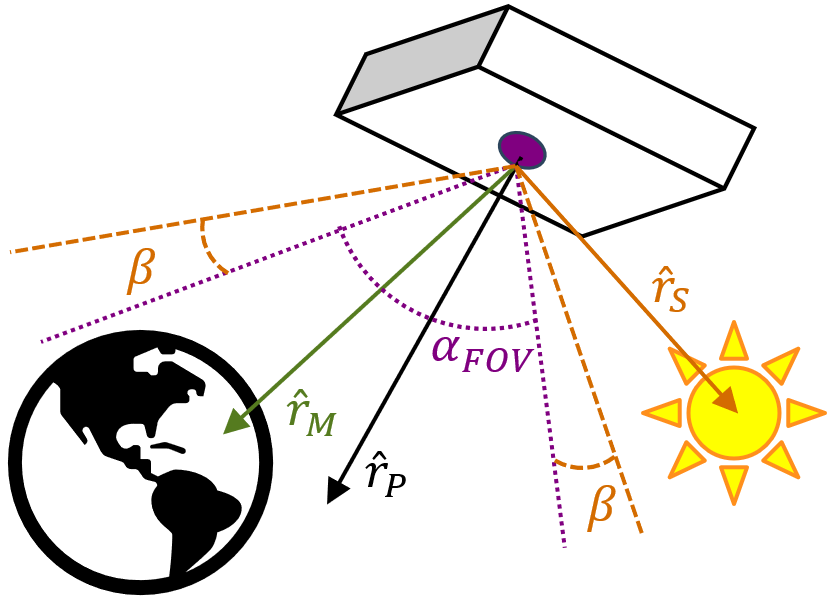}
    \caption{Case study scenario in which the objective is to command the pointing vector $\hat{r}_{P}$ to align with the mission vector $\hat{r}_{M}$, while preventing the Sun vector $\hat{r}_{S}$ from entering the sensor field of view $\alpha_{\text{FOV}}$, with an additional safety buffer $\beta$.}
    \label{fig:Exp3Mission}
\end{figure}

The scenario explored in this case study considers a 6U CubeSat in a circular low-Earth orbit tasked with pointing a body-fixed sensor along a mission-specified direction defined in Hill’s frame, while keeping the sensor outside a Sun-centered exclusion zone, as depicted in Figure \ref{fig:Exp3Mission}. As the Sun moves through the orbital frame, the spacecraft must continuously balance these objectives while minimizing control effort. This case study approaches this scenario by implementing a neural network waypoint generator to produce steering commands that balance mission performance, safety constraint satisfaction, and maneuver efficiency without ground-based planning, as described in greater detail in \cite{santos2026deep}.

\subsection{Verification Process}

In this project, the \ac{NNCS} was first trained using a first-principles, textbook-level model of the space environment described in \cite{santos2026deep}. Using the same first-principles model, combinatorial and adaptive stress testing was conducted to evaluate system performance \cite{behrendt2026combinatorial}. Next, medium-fidelity simulations were run using a model of the flight test vehicle in the open-source Basilisk simulation environment \cite{kenneally2020} to gain additional confidence in performance. 

Testing then progressed to \ac{PIL}, where the Basilisk simulation of the vehicle dynamics ran on a computer connected to the spacecraft processor, as depicted in Figure~\ref{fig:pil}. The autonomy algorithms were placed inside a Docker container on the spacecraft processor and communicated with the simulation computer via \ac{ROS} messages. In \ac{HIL} testing, shown in Figure~\ref{fig:hil}, a twin of the \ac{ACB} from the satellite was connected between the spacecraft processor and the computer system running a spacecraft simulation, so that commands are processed by the spacecraft attitude control subsystem and communicated to the simulation of the spacecraft on the separate terrestrial computer to capture additional hardware integration real-world performance.
 
Emulation testing was not conducted in this case study primarily because the autonomy tests did not include sensing and perception. Following \ac{HIL} testing, the evaluation process progressed to on-orbit testing on a 6U CubeSat. A comparison of results from the first-principles, \ac{PIL}, \ac{HIL}, and on-orbit testing is provided in the next section.

\begin{figure} [htb]
    \centering
    \includegraphics[width=0.5\textwidth]{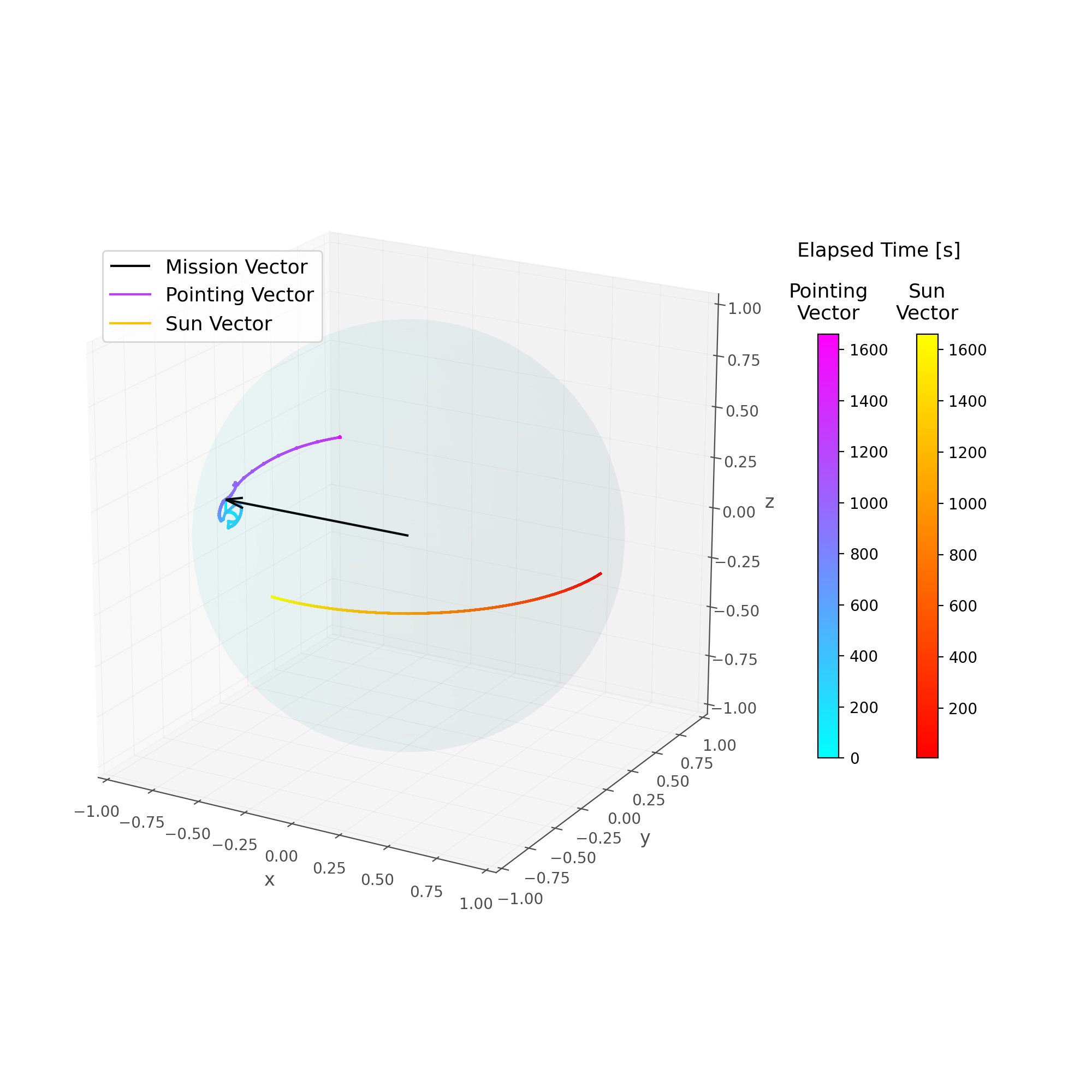}
    \caption{NNCS on-orbit behavior showing achieved pointing $\hat{r}_{P}$, desired mission pointing vector $\hat{r}_{M}$, and Sun location $\hat{r}_{S}$ in Hill's frame.}
    \label{fig:sphere_plot}
\end{figure}

%\begin{figure}[htb]
%    \centering
%    \includegraphics[width=0.5\textwidth]{Figures/keepout_angles_4way_comparison.pdf}
%    \caption{NNCS performance comparison of on-orbit data with \ac{FP} simulation, PIL testing, and HIL testing. (above) Angle between the achieved pointing vector and the desired pointing vector. (below) Angle between the achieved pointing vector and the Sun. }
%    \label{fig:agent_performance}
%\end{figure}

\begin{figure}[htb]
    \centering
    \begin{subfigure}[b]{0.48\textwidth}
        \centering
        \includegraphics[width=\textwidth]{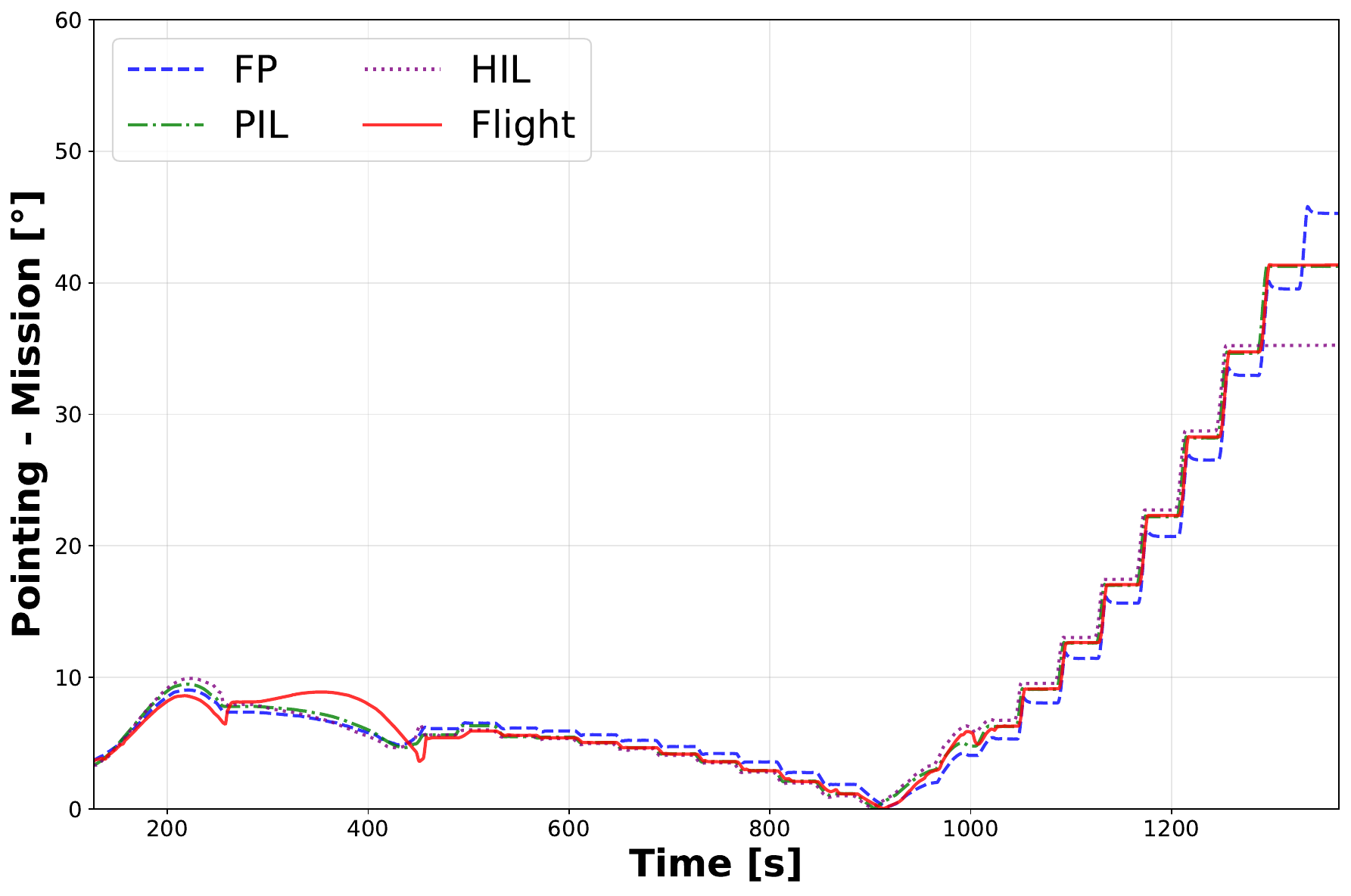}
        \caption{Angle between the achieved pointing vector $\hat{r}_{P}$ and the desired pointing vector $\hat{r}_{M}$.}
        \label{fig:pointing_error}
    \end{subfigure}
    \hfill
    \begin{subfigure}[b]{0.48\textwidth}
        \centering
        \includegraphics[width=\textwidth]{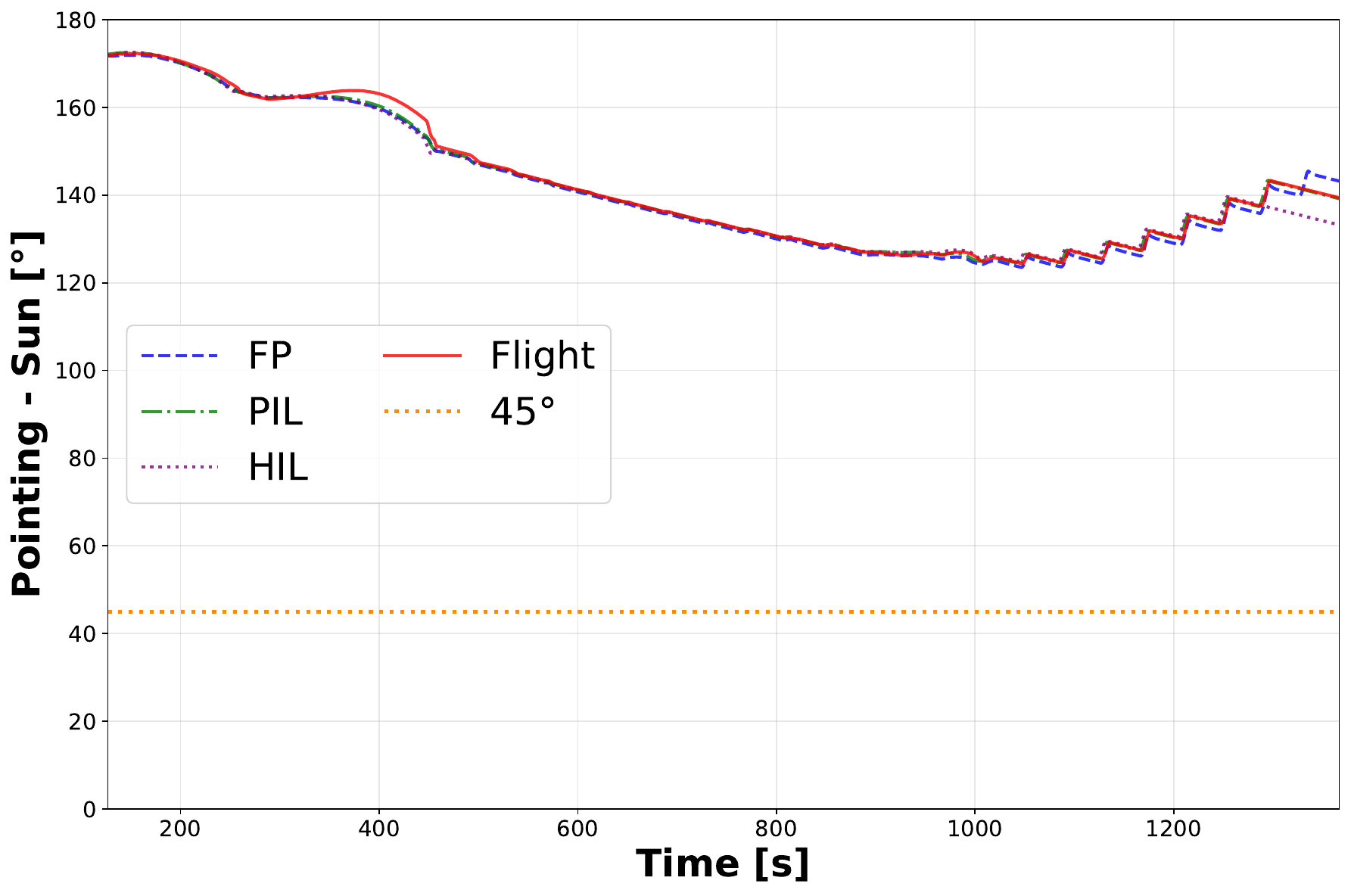}
        \caption{Angle between the achieved pointing vector $\hat{r}_{P}$ and the Sun $\hat{r}_{S}$.}
        \label{fig:sun_margin}
    \end{subfigure}
    \caption{NNCS performance comparison of on-orbit data with \ac{FP} simulation, PIL testing, and HIL testing.}
    \label{fig:agent_performance}
\end{figure}

\subsection{On-Orbit Performance}
During on-orbit testing on a 6U CubeSat, the NNCS took appropriate actions to maintain mission pointing with the body-fixed sensor while ensuring the Sun remained outside the sensor’s exclusion zone. Figure~\ref{fig:sphere_plot} illustrates the neural network’s waypoint-generation behavior throughout the test. The black arrow indicates the mission vector $\hat{r}_{M}$ that the NNCS is attempting to track; the teal-to-magenta line shows the trajectory of the actual pointing vector $\hat{r}_{P}$ over the course of the test; and the red-to-yellow line shows the trajectory of the Sun vector $\hat{r}_{S}$. As the Sun vector approached the mission vector, the NNCS correctly commanded the spacecraft to point away from the mission vector to avoid the Sun and maintain safety.

\subsection{Mission Performance Comparison}
The NNCS exhibited consistent performance across all test environments, as shown in Figures \ref{fig:agent_performance}-\ref{fig:quaternion}, where the behavior in \ac{FP}, \ac{PIL}, and \ac{HIL} test environments is compared to performance in on-orbit flight tests. It is worth noting that exact prediction of flight performance is nearly impossible, as it would require precise knowledge of the vehicle's initial state, a level of precision that is impractical to achieve, especially days or weeks in advance of a flight test. Therefore, pre-flight tests' primary value is in providing trends and expected performance bounds rather than exact predictions. The plots presented in this paper were generated after the flight using the actual initial conditions from the flight test to evaluate how closely the different ground-based simulation environments matched the flight data.

Figure~\ref{fig:agent_performance} presents two key metrics: the mission pointing error (the angle between the mission vector $\hat{r}_{M}$ and the current sensor pointing vector $\hat{r}_{P}$), which is ideally near zero, and the Sun exclusion zone margin (the angle between the sensor pointing vector $\hat{r}_{P}$ and the Sun vector $\hat{r}_{S}$), which is desired to remain above the 45-degree safety threshold. These results demonstrate that the NNCS autonomously balanced task performance (maintaining pointing to the mission vector) with safety constraints (keeping the sensor pointed away from the Sun). Mission pointing accuracy and Sun avoidance behavior from the FP simulation, PIL testing, and HIL testing closely matched the on-orbit results. In all testing venues, the NNCS maintained nominal mission pointing during periods of sufficient Sun separation. As the Sun traversed closer to the exclusion zone, the NNCS commanded attitude waypoints to reorient the sensor, preventing constraint violations while minimizing pointing deviation from mission pointing. The close agreement across testing venues and simulation fidelity levels demonstrates that ground-based testing can accurately predict on-orbit behavior. This shows that the \ac{VandV} Framework can be used to provide confidence in autonomous systems, such as the NNCS, early in the development cycle, prior to flight tests.

\subsection{Simulation Fidelity Verification}
\begin{figure}[htb]
    \centering
    \includegraphics[width=0.5\textwidth]{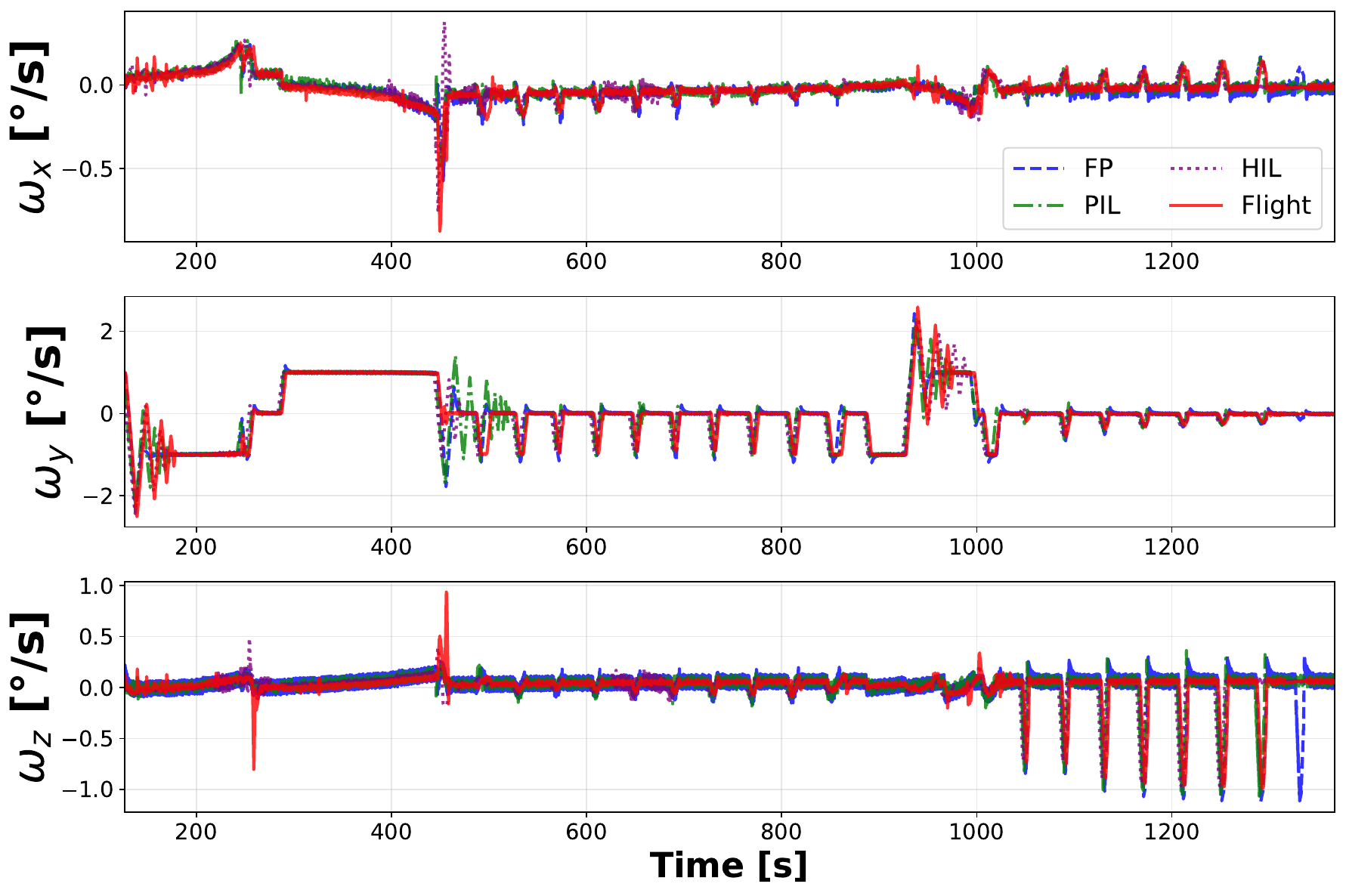}
    \caption{Body angular velocity components~$\omega^B_{BN}$, compared across FP simulation, PIL testing, HIL testing, and flight data.}
    \label{fig:body_rates}
\end{figure}

\begin{figure}[htb]
    \centering
    \includegraphics[width=0.5\textwidth]{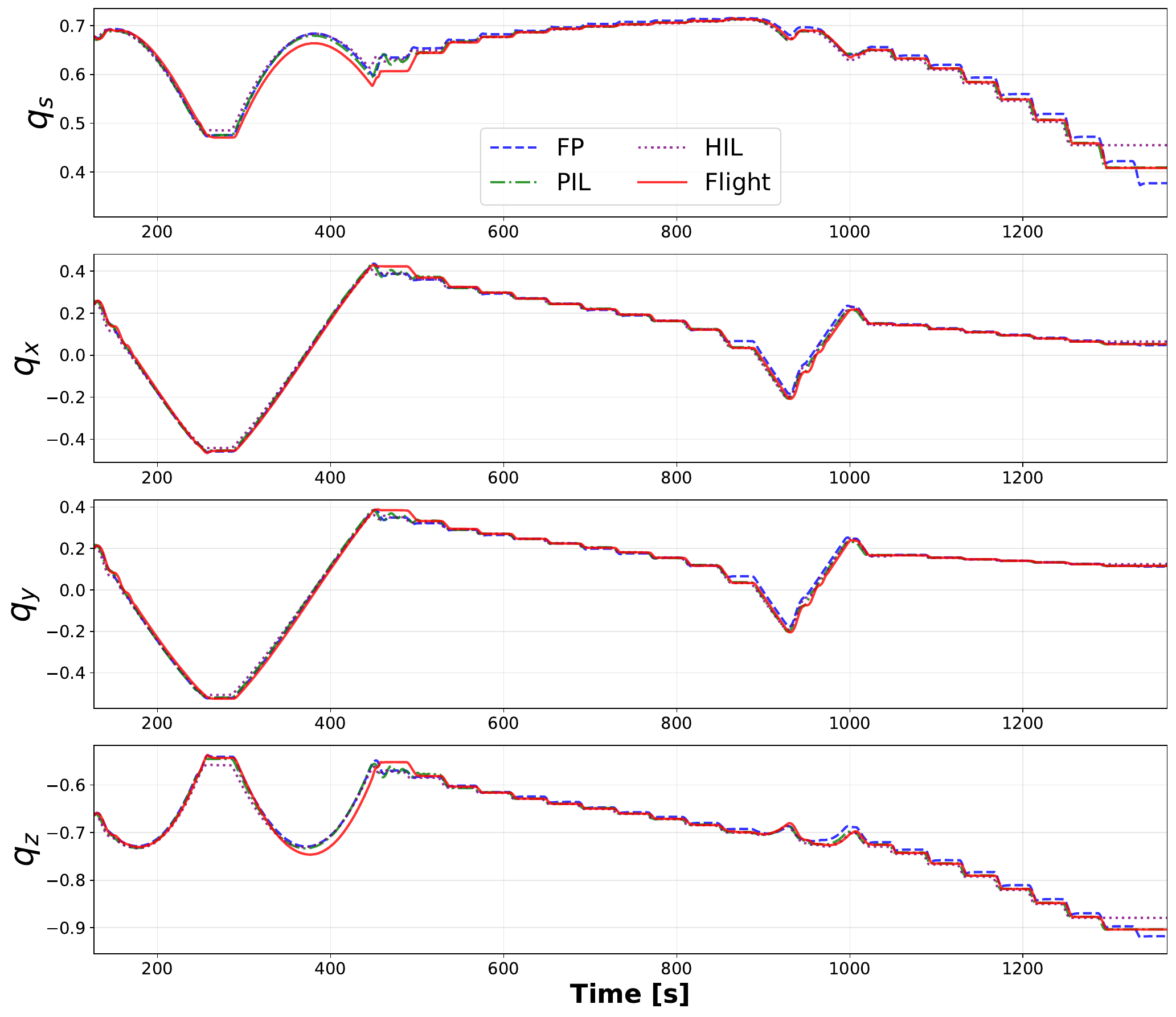}
    \caption{Body-to-Hill frame attitude quaternion components~$q_{BH}$, compared across FP simulation, PIL testing, HIL testing, and flight data.}
    \label{fig:quaternion}
\end{figure}

The spacecraft dynamics observed during on-orbit testing were well-represented by the simulation environments. Figure \ref{fig:quaternion} shows the spacecraft attitude for flight data and simulations from each testing venue, while Figure \ref{fig:body_rates} shows the corresponding body rates during the test. While the attitude error typically remained less than 5 degrees during the majority of the tests, the largest discrepancies occurred during attitude settling following large slew maneuvers, where transient dynamics were dominant. Body rate estimates from the simulations largely matched flight telemetry, with minor differences in settling dynamics and noise characteristics. These results verify that the simulation fidelity in the various test venues provides sufficiently accurate representations of on-orbit dynamics for NNCS evaluation. 

\subsection{Online Verification AI Watchdog Results}
\begin{figure}
    \centering
    \includegraphics[width=\columnwidth]{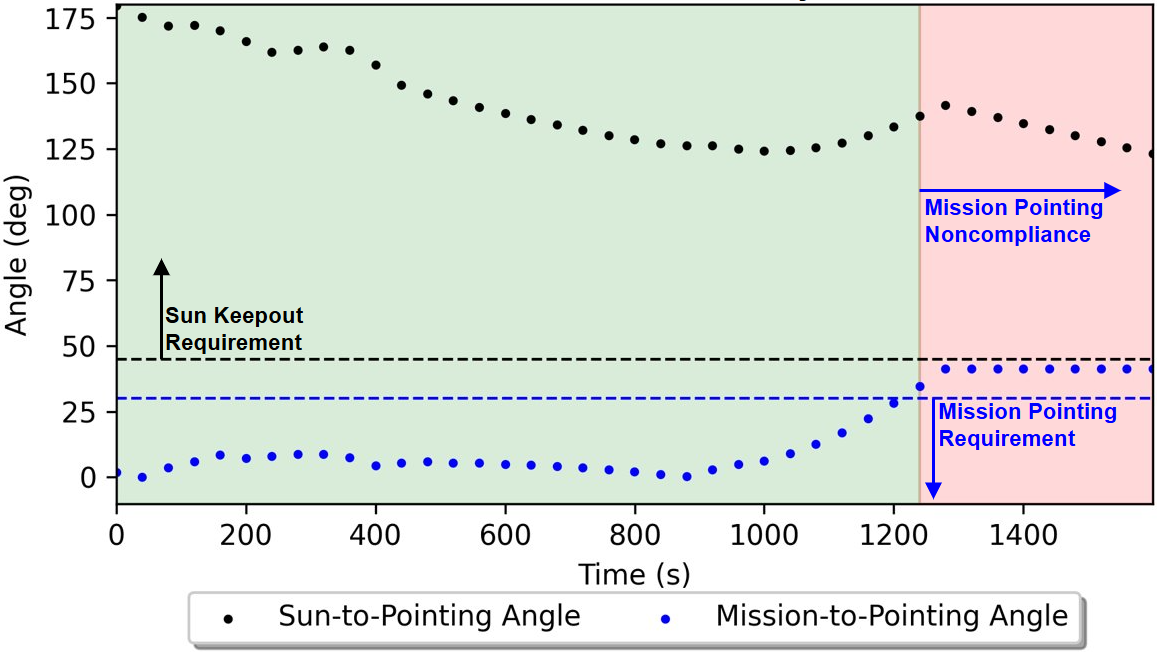}
    \caption{The NNCS was trained to maintain both mission pointing within 30 degrees and sun avoidance by at least 45 degrees. At approximately 1250 seconds, mission pointing is no longer achieved, while the vehicle is maneuvered to avoid violating the sun pointing angle.}
    \label{fig:NNCSPerformance}
\end{figure}

\begin{figure}
    \centering
    \includegraphics[width=\columnwidth]{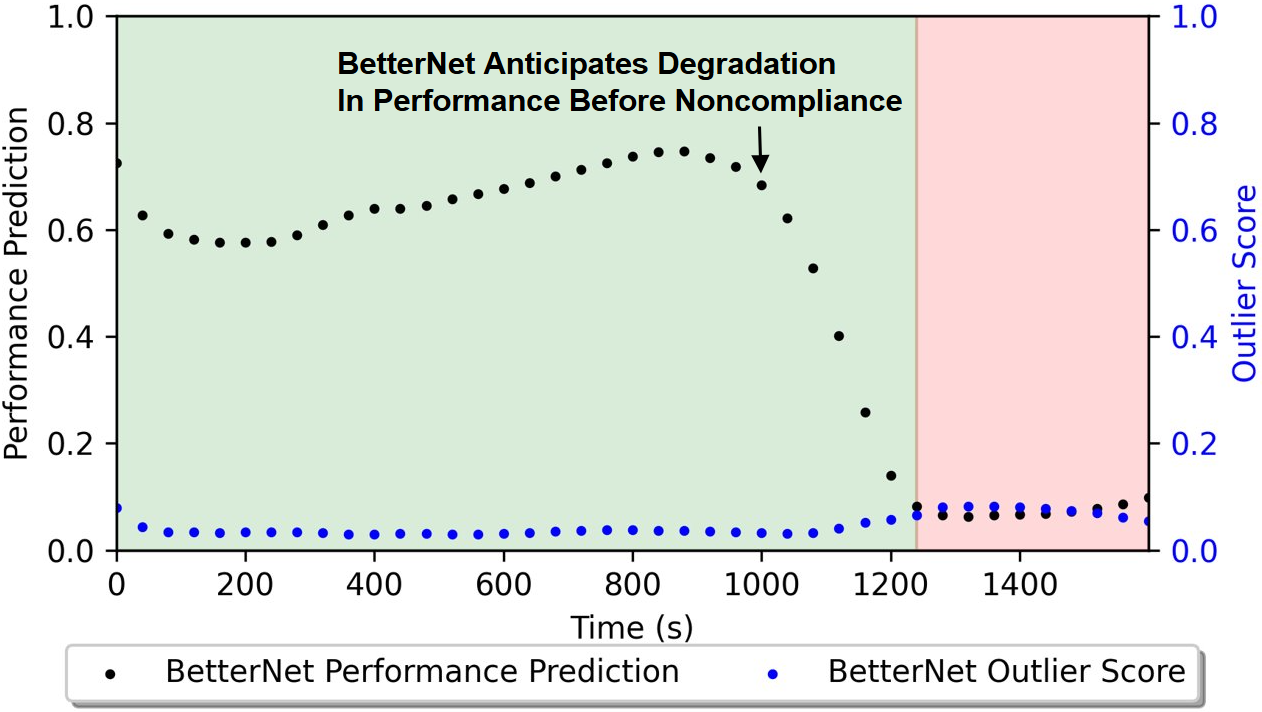}
    \caption{The BetterNet AI Watchdog, trained on medium-fidelity simulation data, is shown to predict NNCS performance degradation in pointing requirements prior to noncompliance. The low outlier score indicates that this scenario is in distribution with the simulated dataset used to fit BetterNet.}
    \label{fig:BetternetPrediction}
\end{figure}

The BetterNet watchdog was trained to predict performance of the \ac{NNCS} in this case study. Training used medium-fidelity simulation results to statistically model the probability of the NNCS achieving both mission pointing and sun avoidance requirements. The model also computes a normalized outlier metric to indicate when BetterNet predictions are being performed outside of the model fitting data distribution. This watchdog model, after being fit, was then evaluated on real-world mission data. Figure \ref{fig:NNCSPerformance} tracks compliance with two mission requirements: 1) bus pointing achieving mission pointing command within 30 degrees and 2) bus pointing avoiding the sun pointing vector by at least 45 degrees. As shown in the figure, the satellite bus maintains sun avoidance requirements but stops meeting mission pointing requirements at approximately 1250 seconds. Figure \ref{fig:BetternetPrediction} demonstrates BetterNet's capability to predict this noncompliance as the predicted performance clearly begins decreasing at approximately 900 seconds. Additionally, the normalized outlier score remains low which indicates that this scenario is in-distribution with scenarios seen when evaluating the NNCS on simulation data.

\subsection{Lessons Learned and Best Practices}

%\note{[Mostly complete - needs review: Key insights from applying the framework. - naming ability to rapidly iterate on the design prior to flight test, since flight test is limited; plus we demonstrated the ability of the framework to provide confidence prior to flight test because flight test results are in margin of error of simulation results; perhaps discussion of what processor, and hardware in the loop got us over baseline]}

The application of the Spacecraft Autonomy \ac{VandV} Framework to this case study demonstrated that several key capabilities proved useful for successful NNCS deployment and provided valuable insights for future testing. 

A primary benefit of the testing framework was the ability to rapidly iterate on the NNCS design prior to on-orbit flight testing, where test opportunities are inherently limited. Prior to flight test, thousands of FP tests, ten PIL tests, and eight HIL tests were conducted, which enabled the identification and resolution of design and integration issues in a controlled ground environment. 

PIL testing revealed the need to decrease commanding frequency to allow adequate spacecraft settling between maneuvers. An error in the calculation of the attitude waypoint commands was also discovered during PIL testing. The framework facilitated iterative collaboration between NNCS developers and the testing team, where feedback from testing was rapidly integrated into subsequent versions and re-evaluated in the test environment. This iterative capability proved invaluable given the limited opportunities for on-orbit testing.

\section{Conclusions and Future Work} \label{sec:conclusions}

This paper presents a comprehensive framework for incremental testing and technology maturation of AI-enabled autonomous spacecraft control systems. The framework integrates offline and online verification approaches to address the critical ``trust gap'' in deploying autonomous systems in the high-consequence space environment.

Key conclusions include:

\begin{itemize}
    \item A multi-phase verification approach with increasing realism and integration provides systematic risk reduction throughout development;
    \item Closed-loop testing at each phase is essential to evaluate autonomous behavior rather than merely system integration;
    \item The combination of AI watchdog systems and safety guardrails provides complementary monitoring and intervention capabilities during flight operations;
    \item The simulation-to-reality transfer challenge can be addressed through systematic progression of offline testing combined with online safety assurance; and
    \item The framework is practical and implementable with mature technologies available today.
\end{itemize}

The path forward for safe deployment of AI-enabled autonomy in spacecraft requires continued development of verification methodologies, standardization of testing approaches, and accumulation of flight heritage. By establishing systematic progression criteria and integrating monitoring with intervention capabilities, this framework provides a viable approach to achieving these goals and supporting the scaling demands of modern space missions.

\subsection{Recommended Testing Strategy}

Based on experience gained during the case study, a streamlined testing strategy has been established for future missions:
\begin{itemize}
    \item Rapid iteration in first-principles simulation for baseline algorithm development and stress testing;
    \item Focused PIL testing to verify timing, interfaces, and processor integration; and
    \item Targeted HIL testing on high-risk scenarios or final flight designs to evaluate robustness to hardware effects.
\end{itemize}
This progression maximizes testing efficiency while maintaining the rigor necessary to predict flight performance with confidence.

Note that a full description and analysis of the \ac{RTA} guardrail performance was too extensive for inclusion here and will be addressed in future work.

\subsection{Future Work: Continuous Test and Evaluation of Autonomy}

A significant open question is how the \ac{VandV} paradigm must evolve as learning shifts from ground-based high-performance computing to advanced in-space processors. Progression will likely occur in three or more phases. The near-term first phase will likely continue the current approach: training only on the ground with onboard inference-only autonomy, allowing ample time to conduct \ac{VandV} pre-deployment. An intermediate second phase may feature continuous training on the ground with new data and periodic updates to the onboard autonomy that runs at inference. Training onboard may become possible in the future with more efficient algorithms or advanced computational hardware. Onboard training by necessity limits the amount of testing that can occur prior to deployment, and safety and mission assurance will rely heavily on watchdogs and guardrails to keep the system within acceptable limits and detect performance degradation. Each of these phases represents a nascent area of active research and development, and concrete \ac{VandV} steps are not yet established. This paper describes the likely path for phases one and two. Verification of continuous learning on orbit is beyond the scope of this work and will require supplementary techniques.

\section*{Acknowledgments}
The authors would also like to acknowledge the contributions of Eddson M. Alcid and John F. Swigart to the development of the HIL testbed described in this project. 

In accordance with transparency practices for the use of artificial intelligence in technical documentation, the authors disclose that generative \ac{AI} tools were employed to assist with editing and improving the clarity of written content in this report. These tools were used solely for language refinement and document formatting purposes. All substantive technical content, including analysis, methodology, findings, conclusions, and recommendations, represents the original intellectual work of the authors. The authors have reviewed all \ac{AI}-assisted edits for accuracy and appropriateness. The authors assume full responsibility for the technical accuracy, validity, and integrity of all information, data, and conclusions presented in this document.

The views expressed are those of the authors and do not reflect the official guidance or position of the United States Government, the Department of War, or of the United States Air Force. Approved for public release; distribution unlimited. Case Number AFRL-2026-3817.

\section*{Abbreviations}
% abbrevs.tex - Spacecraft Autonomy V&V Paper
\begin{acronym}
    \acro{ACB}{Attitude Control Board}
    \acro{AFRL}{Air Force Research Laboratory}
    \acro{AI}{Artificial Intelligence}
    \acro{CAVE}{Collaborative Autonomous Vehicles Ecosystem}
    \acro{FP}{First Principles}
    %\acro{ESA}{European Space Agency}
    %\acro{FDIR}{Fault Detection, Isolation, and Response}
    %\acro{GPS}{Global Positioning System}
    %\acro{HAT}{Human-Autonomy Teaming}
    \acro{HIL}{Hardware-in-the-Loop}
    \acro{HITL}{Human-in-the-Loop}
    %\acro{ISAM}{In-Space Servicing, Assembly, and Manufacturing}
    %\acro{LEO}{Low Earth Orbit}
    %\acro{ML}{Machine Learning}
    %\acro{NASA}{National Aeronautics and Space Administration}
    \acro{NNCS}{Neural Network Control System}
    %\acro{OODA}{Observe-Orient-Decide-Act}
    \acro{PIL}{Processor-in-the-Loop}
    \acro{PPO}{Proximal Policy Optimization}
    \acro{RL}{Reinforcement Learning}
    \acro{ROS}{Robot Operating System}
    \acro{RTA}{Runtime Assurance}
    \acro{RWA}{Reaction Wheel Array}
    \acro{STARS}{Safe Trusted Autonomy for Responsible Space}
    %\acro{TandE}[T\&E]{Test and Evaluation}
    \acro{TRLs}{Technology Readiness Levels}
    \acro{VandV}[V\&V]{Verification and Validation}
\end{acronym}
\bibliographystyle{IEEEtran}
\bibliography{references}

% If you don't have a .bib file yet, you can use:
% \begin{thebibliography}{99}
% \bibitem{reference_label}
% Author names, ``Title,'' \textit{Journal/Conference}, vol. X, no. Y, pp. Z, Year.
% \end{thebibliography}

%%%%%%%%%%%%%%%%%%%%%%%%%%%%%%%%%%%%%%%%%%%%%%%%%%%%%%%%%%%%%%%%%%%%%%%%%%%%%%%%%%%%%%%%%%%%%%%%%%%%%%
\thebiography
%% This biostyle allows you to insert your photo size 1in X 1.25in
\begin{biographywithpic}
{Dr. Kerianne Hobbs}{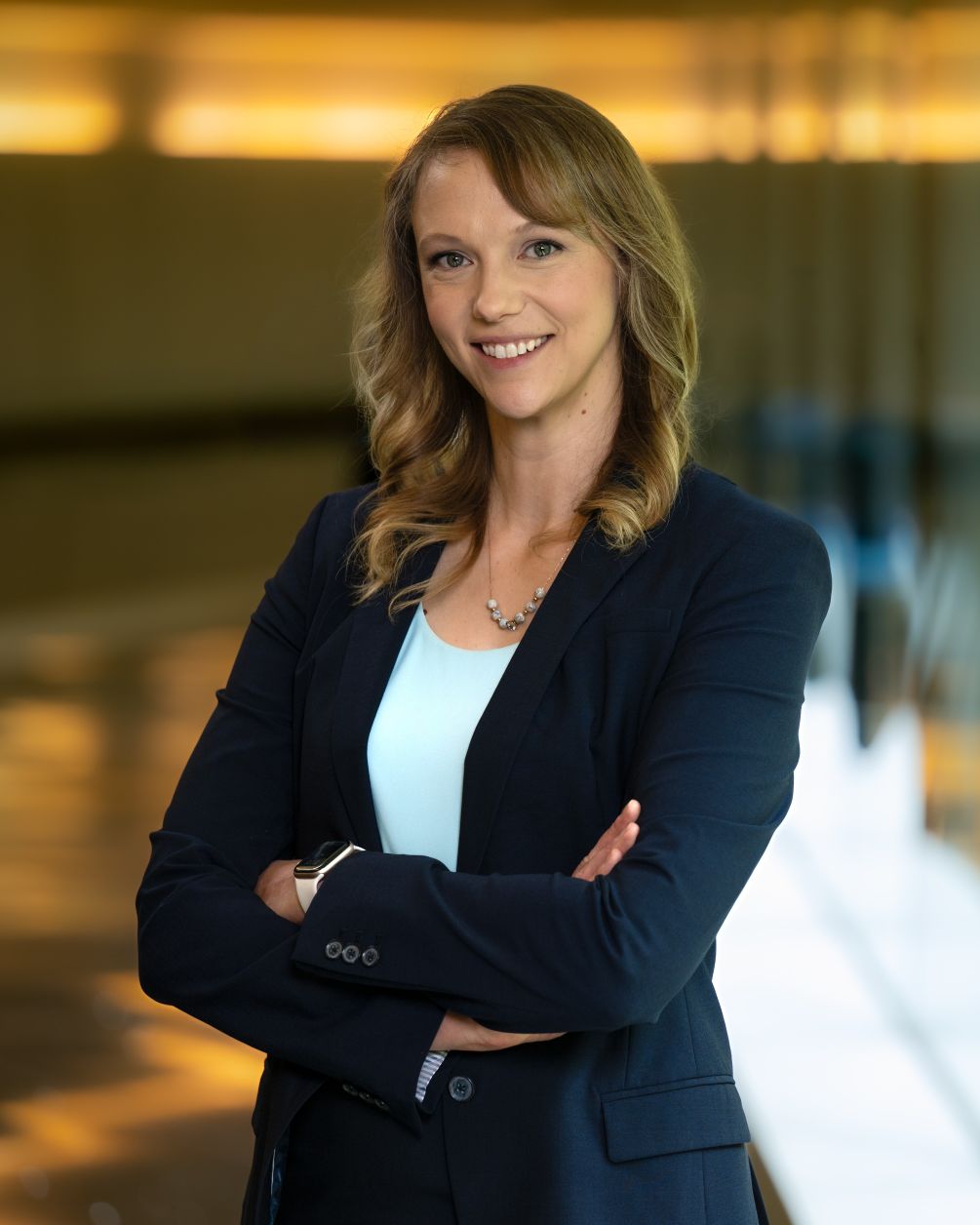}
leads space vehicle autonomy and trust research at The Aerospace Corporation. Previously, as Safe Autonomy Lead at the Air Force Research Laboratory, she directed the Safe Trusted Autonomy for Responsible Spacecraft (STARS) program, developing neural network control systems using reinforcement learning, runtime assurance technologies, and human-AI interfaces for multi-vehicle autonomous satellite operations. Her research expertise includes formal methods and runtime assurance for autonomous software assurance. Prior to STARS, she contributed to the development and testing of automatic collision avoidance technologies for the F-16. She holds a B.S. in Aerospace Engineering from Embry-Riddle Aeronautical University, an M.S. in Astronautical Engineering from the Air Force Institute of Technology, and a Ph.D. in Aerospace Engineering from Georgia Institute of Technology. Dr. Hobbs was a member of the inaugural AIAA 30/30 class in 2026, is a 2024 AIAA Associate Fellow and 2020 AFCEA 40 Under 40 recipient. She was part of the team awarded the 2018 Collier Trophy for the Automatic Ground Collision Avoidance System.
\end{biographywithpic} 

\begin{biographywithpic}{Ben Bycroft}{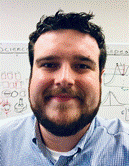}~is Director of Vehicle Autonomy \& System Trust at The Aerospace Corporation. His work spans research, development, and application of advanced and AI-enabled autonomy, dynamics and control, as well as electromechanical systems engineering for both launch vehicles and spacecraft. Benjamen has previously taught courses within the University of Southern California’s Viterbi School of Engineering. He holds a B.S. in Mechanical Engineering and an M.S. in Dynamics \& Control from USC, as well as a B.A. in Physics from Willamette University.
\end{biographywithpic} 

\begin{biographywithpic}{Pamela S Wood}{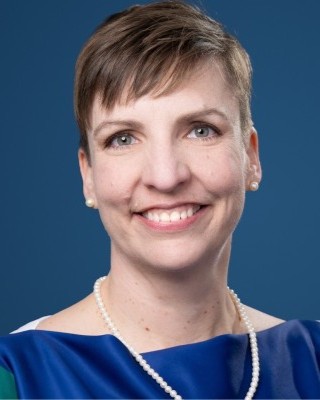}
is a Senior Vehicle Engineering and Autonomy Project Leader at The Aerospace Corporation. Her work focuses on integration and assessment of artificial intelligence for advanced autonomy in space systems, as well as assurance methods for cognitive adaptive systems. Ms. Wood has a background data analysis and exploitation, with a B.S. in Industrial Systems Engineering from Virginia Tech, as well as an M.S. in Data Analytics Engineering from George Mason University where she is also pursuing a Ph.D. in Computational Sciences and Informatics.
\end{biographywithpic} 

\begin{biographywithpic}{Mathew W. Bissonnette}{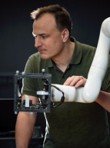}~is Senior Engineering Specialist and Collaborative and Autonomous Vehicle Ecosystem Lab Manager at The Aerospace Corporation, leading modeling and simulation for vehicle autonomy systems. His expertise includes control systems design, embedded software, and hardware-in-the-loop testing. Previously, he led flight controls analysis for Atlas V and Falcon 9 launch vehicles and served as Principal Software Engineer at Northrop Grumman. He holds an M.S. in Dynamic Systems and Controls from UT Austin (2014) and a B.S. in Mechanical Engineering from Cal Poly (2012).
\end{biographywithpic} 

\begin{biographywithpic}{Michael J. Durst}{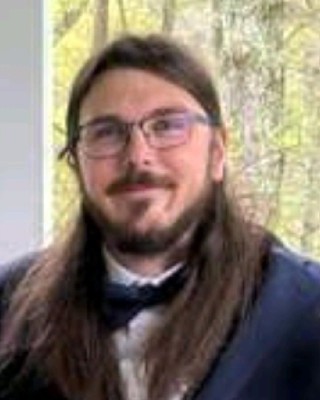} is Manager of the Advanced and Distributed Autonomy section at The Aerospace Corporation, leading modeling and simulation for autonomy verification and validation testbeds. His work focuses on verification of reinforcement learning for spacecraft autonomy. Previously, he was Senior Member of Technical Staff at Aerospace and spent four years as a systems engineer at TMC Technologies supporting NASA's Katherine Johnson Independent Verification and Validation (IV\&V) Facility. He holds a B.S. in Aerospace Engineering from West Virginia University (2019).
\end{biographywithpic} 

\begin{biographywithpic}{Kimberly Sover}{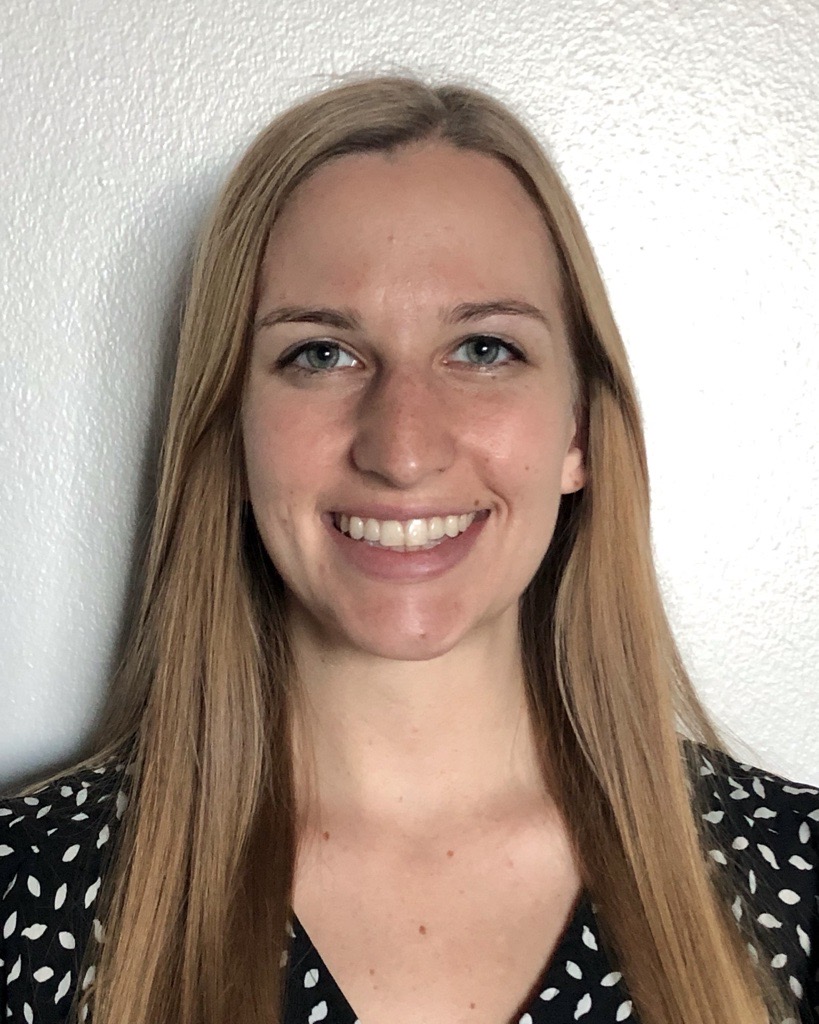}~is an Engineering Manager in the Vehicle Autonomy and System Trust Department. Her work focuses on integrated test and analysis of autonomous systems, with a background in guidance, navigation, and control including actuators, sensors, and software for spacecraft and launch vehicles. She holds an M.S. in Mechanical Engineering from UC San Diego and a B.S. in Mechanical Engineering from UC Berkeley.
\end{biographywithpic}

\begin{biographywithpic}{Alexandra Mangel}{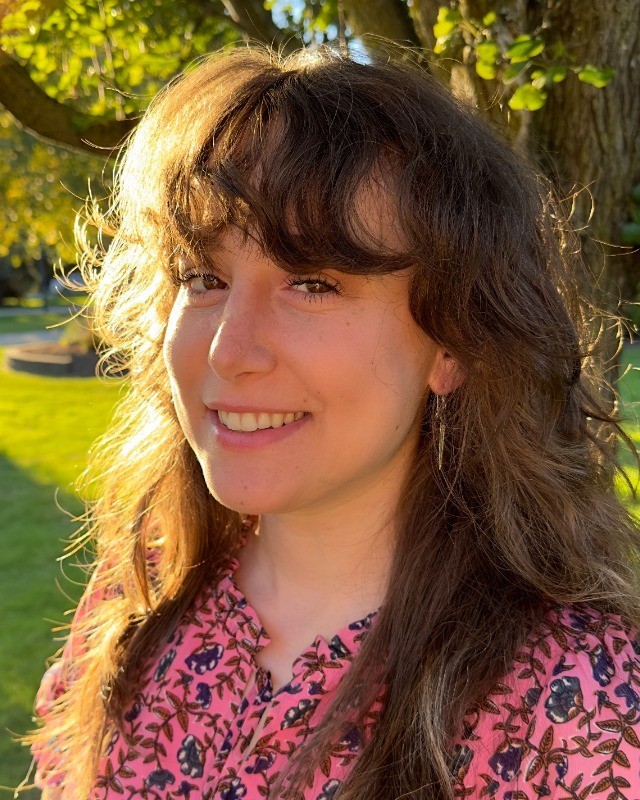}~is a Senior Member of Technical Staff in the Vehicle Autonomy \& System Trust department at the Aerospace Corporation. Her work pertains to the incorporation of autonomy for spacecraft control applications and developing verification and validation processes to enable trustworthiness in AI/ML algorithms. She has a B.S. in Aerospace Engineering from the Ohio State University and an M.S. in Aerospace Engineering from the University of Maryland.
\end{biographywithpic} 

\begin{biographywithpic}{Dr. Gabriel Behrendt}{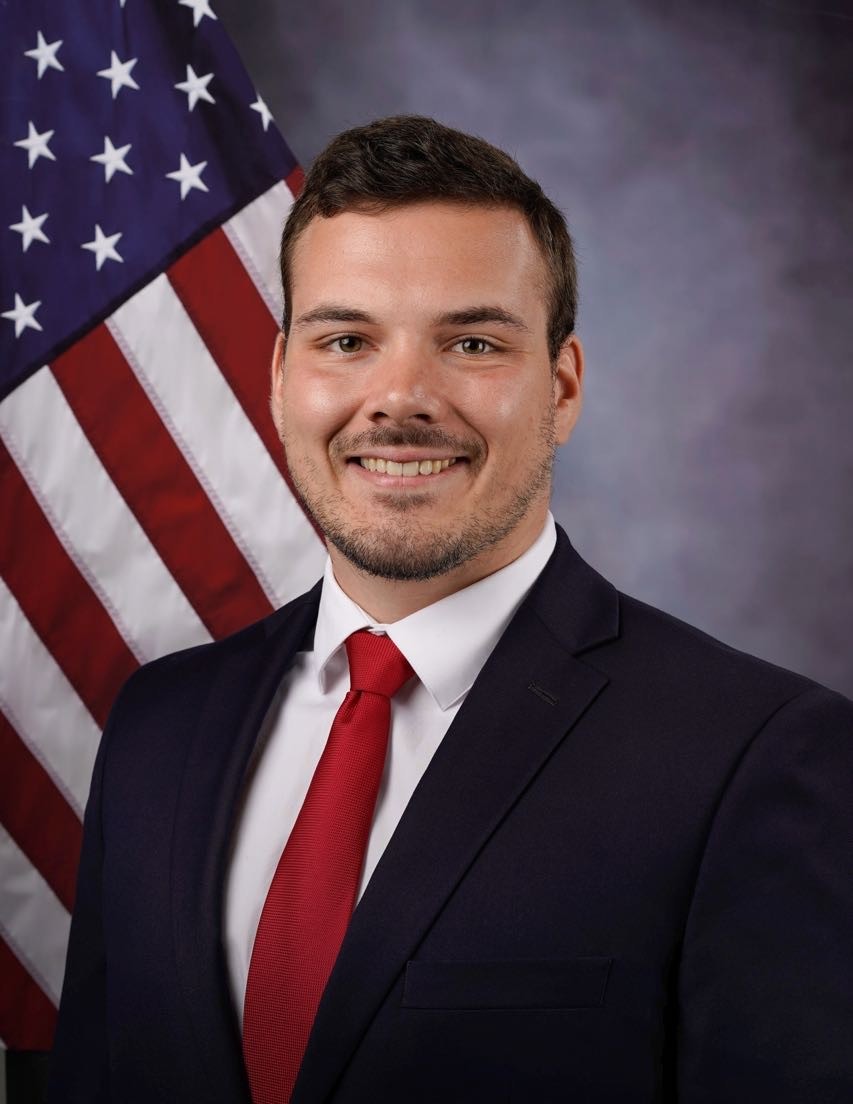}~is a Senior Member of Technical Staff in the Vehicle Autonomy and System Trust department at The Aerospace Corporation. Gabriel holds a B.S. in Mechanical Engineering from the University of Iowa, an M.S. in Aerospace Engineering from the University of Florida, and a Ph.D. in Electrical and Computer Engineering from Georgia Tech. His work focuses on developing intelligent testing and evaluation methods for autonomous aerospace systems, as well as modeling and simulation for advanced satellite guidance, navigation, and control. 
\end{biographywithpic}

\begin{biographywithpic}{Upneet Singh}{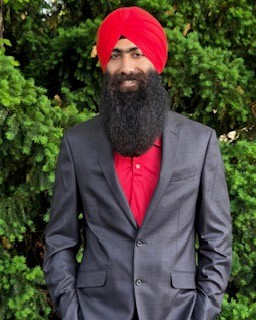} 
 is a Member of Technical Staff in the Vehicle Autonomy and System Trust department at Aerospace Corporation. He holds a Bachelor's in Computer Engineering and Master's degree in Electrical Engineering from George Mason University, focused in robotics and control. His focus at Aerospace is in developing simulation testbeds used to qualify and test autonomous space capabilities. He also has a long history with embedded systems and robotics GNC algorithms, which contributes to his work developing new capabilities in the Collaborative Autonomous Vehicle Ecosystem (CAVE) lab.
\end{biographywithpic} 

\begin{biographywithpic}
{Juan Rodriguez}{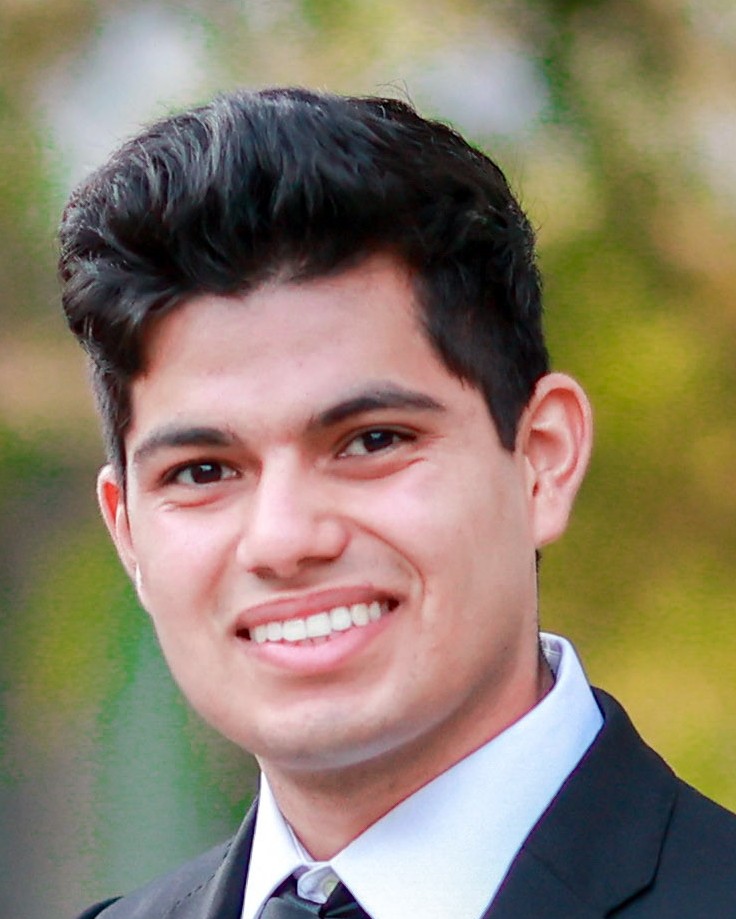} is a Member of Technical Staff in the Vehicle Autonomy \& System Trust department at The Aerospace Corporation. His focus is on the development of tests and testbeds to build trust in autonomous systems. He also has experience with robotics while working in the Collaborative Autonomous Vehicle Ecosystem (CAVE) Lab. He graduated from California State University, Long Beach with a B.S in aerospace engineering with a focus on astronautics.
\end{biographywithpic} 

\begin{biographywithpic}{Lukas Dillingham}{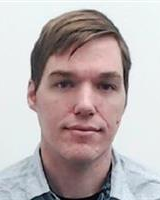}~is an Engineering Specialist in Vehicle Systems Software Department at the Aerospace Corporation. Lukas has a B.S. in Physics from Northern Arizona University, and a M.S. in Computer Engineering from Rochester University. He primarily works on hardware-in-the-loop (HIL) testbed setup and testing for launch vehicles, as well as flight software development for Aerospace's small satellite program Aerocube.
\end{biographywithpic} 

\begin{biographywithpic}
{Zeve Akerling}{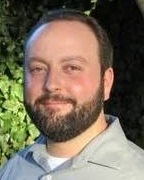} is an engineer at The Aerospace Corporation. His previous experience includes work in quantum computing, writing a compiler for a programmable quantum annealer at USC; work on superconducting electronics at NASA JPL; and computational chemistry simulation for pharmaceutical drug design and molecular nanotechnology at UCLA. He holds a double B.S. in Physics and Chemistry from the University of California, Los Angeles, and a M.S. in Physics from the University of Southern California. 
\end{biographywithpic} 

\begin{biographywithpic}{Nicholas Oune}
{Figures/Bios/NickOune.png} is an Engineering Manager in the Control Analysis Department at The Aerospace Corporation. His work focuses on payload pointing estimation and calibration, navigation, satellite attitude control system (ACS) simulation development. Nick has a background in control design and analysis, state and parameter estimation, and stochastic processes. He has a B.S. and M.S. in Mechanical Engineering from UC Irvine.  

\end{biographywithpic} 

\begin{biographywithpic}{Daniel Thomlinson}{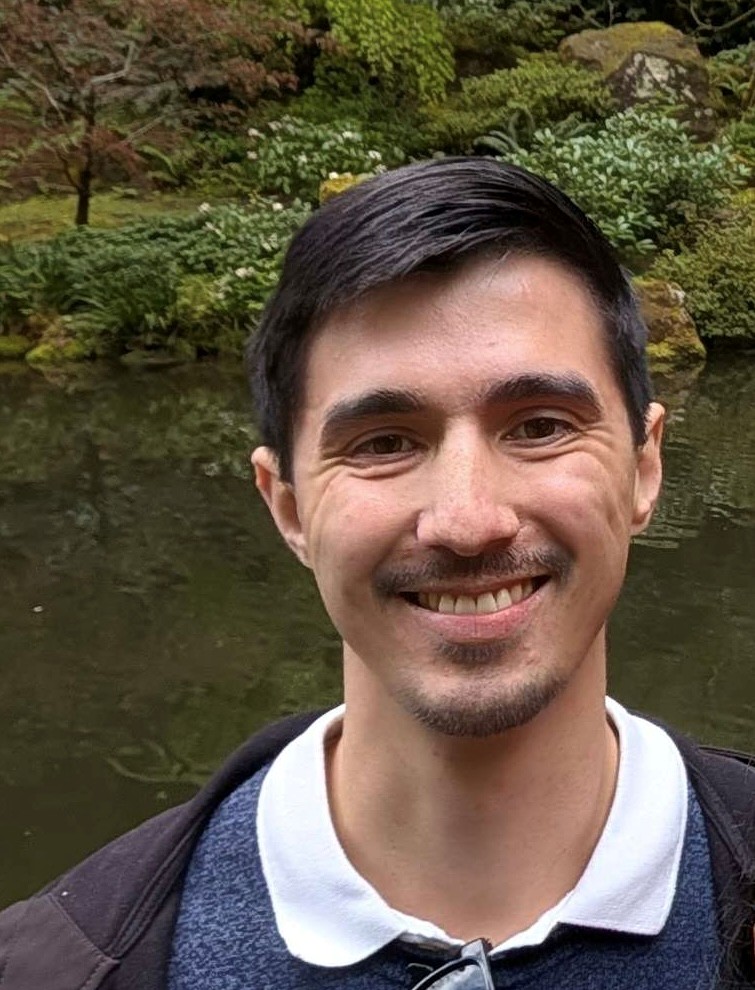} is an Engineering Specialist in the Control Analysis Department at The Aerospace Corporation. His work focuses on simulation development, state estimation, dynamics modeling, and linear covariance analysis for satellite systems. Daniel has a background focused on attitude control and determination and rendezvous proximity operations, with a B.S. in mechanical engineering from UCLA and an M.S. in Aeronautics and Astronautics from Stanford University.
\end{biographywithpic} 

\begin{biographywithpic}{Dr. Evan Sperber}{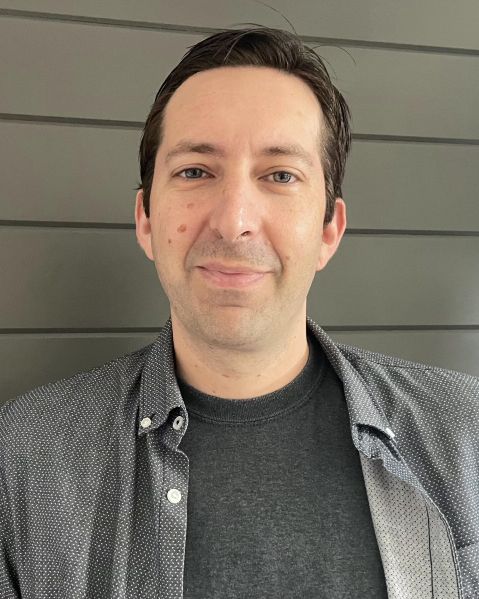} is a Senior Engineering Specialist in the Control Analysis Department at  the Aerospace Corporation. His work focuses on multibody dynamics with specialization in designing and implementing simulations for spacecraft attitude dynamics and control. Dr. Sperber holds a B.A. in Physics from UC Berkeley and a Ph.D. in Mechanical and Aerospace Engineering from UC Davis with a concentration in spacecraft attitude dynamics.
\end{biographywithpic} 

\begin{biographywithpic}
{Denise Burrell}{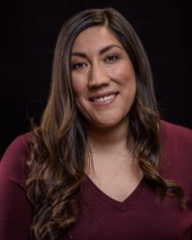} is a Senior Engineering Specialist in the Control Analysis Department at the Aerospace Corporation. Her work focuses on spacecraft attitude control modeling and the development of integrated simulation environments that interface multiple software platforms. Denise holds a B.S. in Aerospace Engineering from Cal Poly Pomona and an M.S. in Dynamics \& Control from USC, with specialization in modeling and simulation of attitude control and determination systems and rendezvous and proximity operations.
\end{biographywithpic} 

\begin{biographywithpic}
{Phaedrus Leeds}{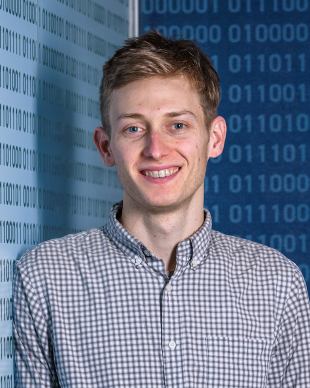} earned his B.S. in Computer Science from The University of Alabama in 2017. As a software engineer at Endless, a nonprofit expanding access to educational technology in the Global South, he led the development of offline updates technology and contributed extensively to open-source software across the Linux Desktop stack. Currently at The Aerospace Corporation's Small Satellite Department, Phaedrus leads edge computing initiatives for space applications. Phaedrus collaborates with teams across government, industry, and academia to enable flight demonstrations of their technologies on CubeSat platforms.
\end{biographywithpic} 

\begin{biographywithpic}
{Dr. Brian Hardy}{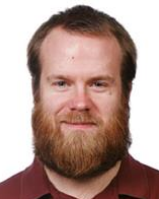}~is a Principal Engineer in The Aerospace Corporation’s Small Satellite Department. Dr. Hardy joined The Aerospace Corporation in 2001 and has designed, performed, and analyzed data from a variety of fluid mechanics, structural mechanics, and thermal experiments, to develop new technologies as well as evaluate reliability and diagnose failures of existing spacecraft and launch systems. Dr. Hardy has been working for the past fifteen years to build, calibrate, test, and qualify sensors and subsystems for small satellites built by The Aerospace Corporation, and is presently the lead engineer in charge of integration, test and on-orbit operations. He holds a B.S. in Mechanical Engineering from Caltech and a Ph.D. in Aerospace Engineering from UCLA.
\end{biographywithpic} 

\begin{biographywithpic}{Dr. Nathaniel ``Nate" Hamilton}{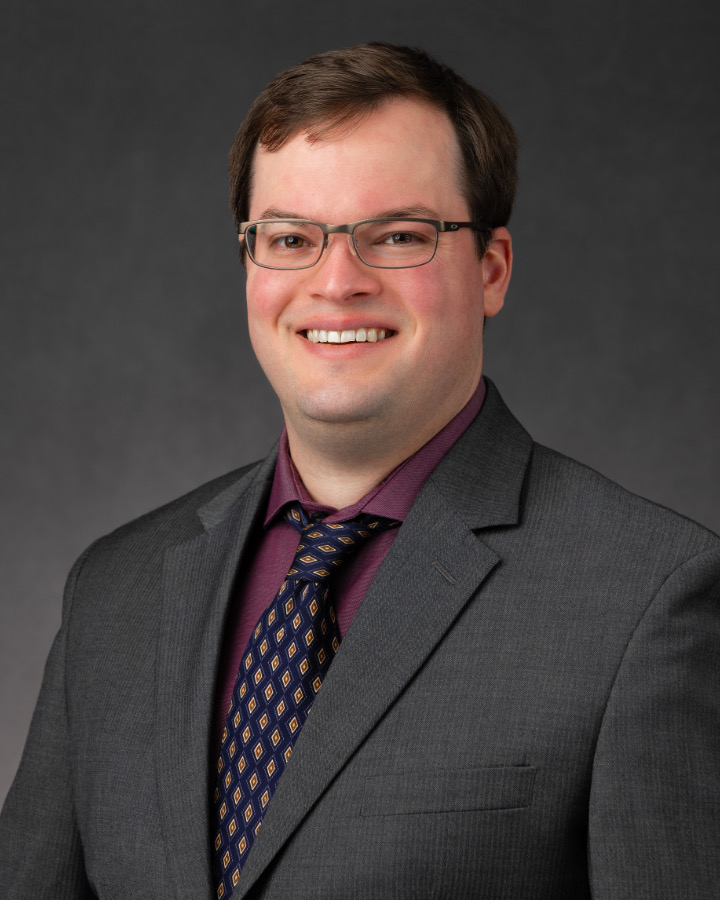} is the Team Lead and Reinforcement Learning Engineer for University of Dayton Research Institute’s Space Autonomy Team, which primarily supports the Air Force Research Laboratory's Autonomy Capability Team (ACT3). There he investigates Safe Reinforcement Learning (SafeRL) approaches for integrating safety into the learning process to enable safe, trusted, and certifiable autonomous and learning-enabled controllers for aircraft and spacecraft applications, and integrates those solutions on real world platforms. His previous experience includes studying how Run Time Assurance (RTA) impacts the learning and performance of SafeRL agents, and work in simulation to real-world (sim2real) transfer for learning-enabled controllers. In 2019, Dr. Hamilton was awarded the National Defense Science and Engineering Graduate (NDSEG) Fellowship. Dr. Hamilton has a B.S.in Electrical and Computer Engineering from Lipscomb University, and an M.S. and Ph.D. in Electrical Engineering from Vanderbilt University.
\end{biographywithpic}

\begin{biographywithpic} 
{Raul Santos}{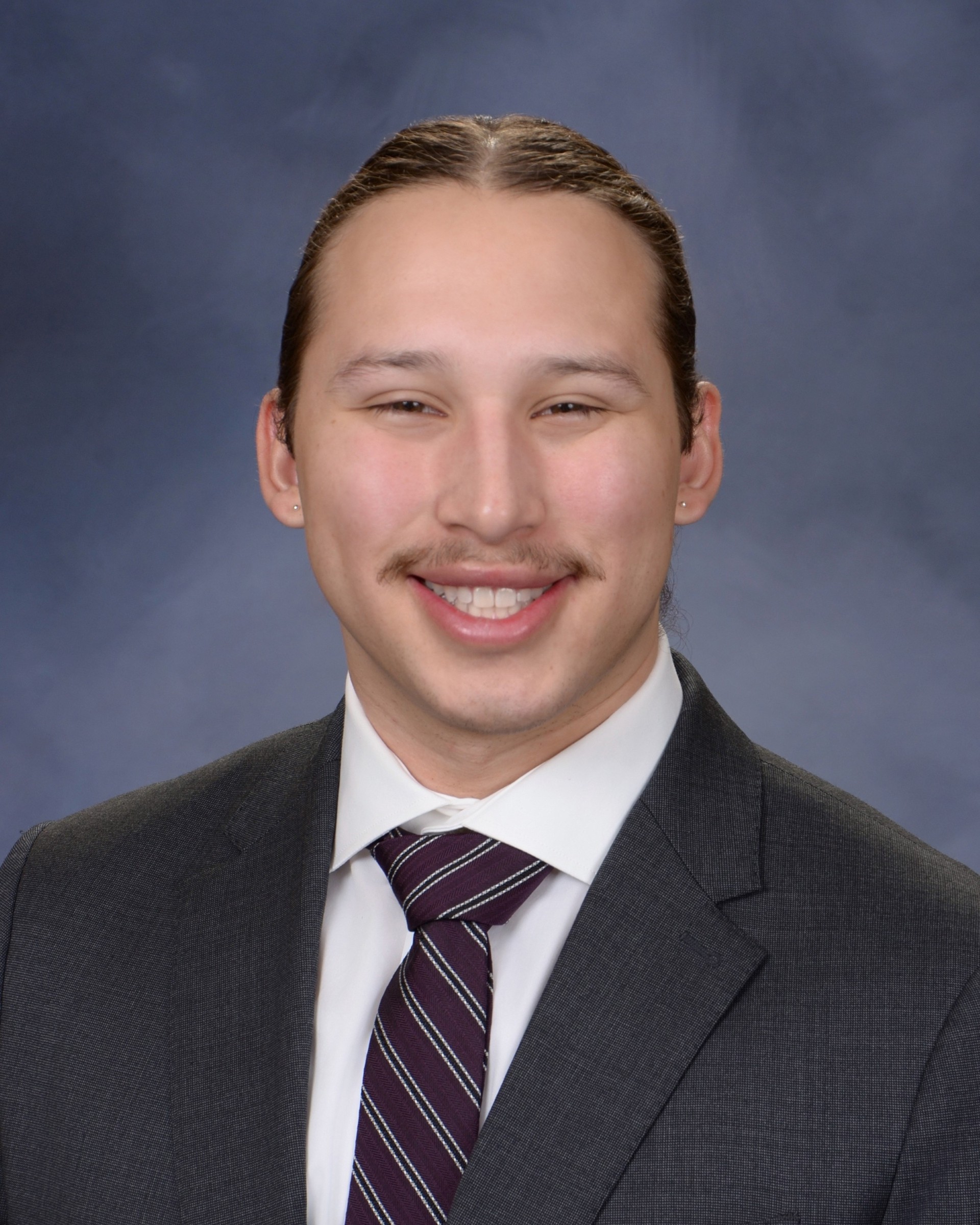} is an Aerospace Engineering M.S. student at Texas A\&M University and a Graduate Research Assistant in the Vehicle Systems \& Control Laboratory. His research focuses on guidance, navigation, and control, autonomous systems, and learning-enabled autonomy for spacecraft rendezvous and proximity operations. He also works as a Safe Autonomy Engineering Intern with the University of Dayton Research Institute, supporting the Air Force Research Laboratory’s Safe Autonomy Team, where his work has included reinforcement learning-based spacecraft guidance, uncertainty-aware autonomous search and tracking, and the transition of learning-enabled control algorithms from simulation to flight testing. He received his B.S. in Aerospace Engineering from Texas A\&M University in 2025.
\end{biographywithpic} 

\begin{biographywithpic}
{Arturo de la Barcena}{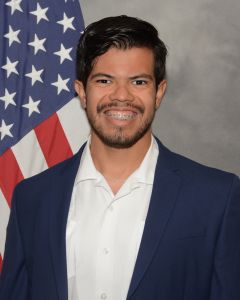} is a Safe Autonomy Engineer with the University of Dayton Research Institute, supporting the Autonomy Capability Team (ACT3) at the Air Force Research Laboratory. His work focuses on safety assurance for autonomous systems, with research interests in air and space vehicle control, autonomy, and safety assurance. His previous research includes quadrotor flight control, terrestrial surrogate testing for space vehicles, and safety assurance techniques for air and space applications. He received his B.S. and M.S. degrees in Mechanical Engineering from the University of Houston.
\end{biographywithpic} 

 \begin{biographywithpic}
{Kyle Dunlap }{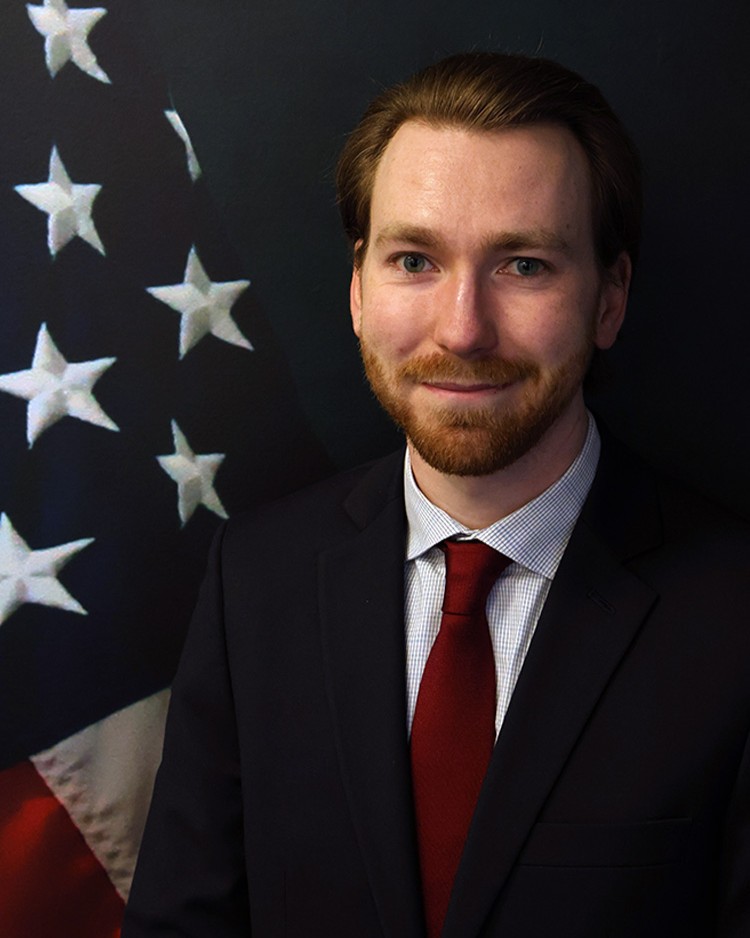}
is the Safe Autonomy and Space Lead on the Autonomy Capability Team (ACT3) at the Air Force Research Laboratory. There he investigates safe, trusted, ethical, and certifiable autonomous and learning controllers for spacecraft applications. His previous experience includes developing a universal framework for run time assurance and comparing different safety assurance approaches during reinforcement learning training. He received his B.S. and M.S.in Aerospace Engineering from the University of Cincinnati.
\end{biographywithpic}

%\begin{biographywithpic}[{\includegraphics[width=1in,height=1.25in,clip,keepaspectratio]{figs/bios/darren.jpg}}]{Darren W. Rowen}~ is Principal Engineer for Applied Edge Computing at The Aerospace Corporation's xLab, where he guides integration of edge computing with prototypes and spearheads the design of satellite and payload architectures to support autonomous operations. He earned his combined B.S./M.S. in Electrical Engineering from the Rochester Institute of Technology, focusing on control systems. Mr. Rowen joined Aerospace in 2005 as a Member of the Technical Staff specializing in control systems, and after working at Northrop Grumman Space Technology (2007-2011) supporting control systems analysis for Class A spacecraft, he rejoined Aerospace in 2011. He became the lead developer for one of the first CubeSat-size attitude determination and control systems and was instrumental in demonstrating the first space-to-ground optical communication link from a CubeSat. He served as Manager of the Small Satellite Development \& Operations section (2014) and Director of the Small Satellite Department (2018-2025), enabling the successful launch and operations of more than 25 custom small satellites forming one of the biggest privately managed satellite constellations.
%\end{biographywithpic} 

\end{document}